# Plane wave in a moving medium and resolution of the Abraham-Minkowski debate by the special principle of relativity

Changbiao Wang (changbiao_wang@yahoo.com)
ShangGang Group, 70 Huntington Road, Apartment 11, New Haven, CT 06512, USA

In this paper, a novel approach for resolution of the Abraham-Minkowski debate is proposed, in which the principle of relativity is used to uniquely determine the light momentum formulation for a plane wave in a moving non-dispersive lossless isotropic uniform medium. It is shown by analysis of the plane-wave solution that, (1) there may be a pseudo-power flow when a medium moves, and the Poynting vector does not necessarily denote the direction of real power flowing, (2) Minkowski's light momentum and energy constitute a Lorentz four-vector in a form of single photon or single EM-field cell, and Planck constant is a Lorentz invariant, (3) there is no momentum transfer taking place between the plane wave and the uniform medium, and the EM momentum conservation equation cannot be uniquely determined without resort to the principle of relativity, and (4) the moving medium behaves as a so-called "negative index medium" when it moves opposite to the wave vector at a faster-than-dielectric light speed. It is also shown by analysis of EM-field Lorentz transformations that, when a static electric (magnetic) field moves in free space, neither Abraham's nor Minkowski's formulation can correctly describe a real electromagnetic momentum; as an application of this principle, the classical electron mass-energy paradox is analyzed and resolved. Finally, a general EM momentum definition is proposed, and according to this new definition, the traditional "Abraham-type" and "Minkowski-type" momentums in the dispersion wave-guiding systems, such as regular dielectric-filled metallic waveguides, are found to be included in the same momentum formulation, but they appear at different frequencies.

## I. INTRODUCTION

The momentum of light in a medium is a fundamental question [1-40]. Maxwell equations support various forms of momentum conservation equations [4,5,11,18,23]. In a sense, the construction of electromagnetic (EM) momentum is artificial or even "arbitrary" [11], which is a kind of indeterminacy. However it is the indeterminacy that results in the question of light momentum. For example, in a recent Letter by Barnett [19], a total-momentum model is proposed to support both Minkowski's and Abraham's formulations of light momentum: one is canonical, and the other is kinetic, and they are both correct [20].

Clearly, it is an insufficiency of existing theories [11,19,22] that the light momentum cannot be uniquely determined. For example, according to the existing theories the momentum of a specific photon in a medium, observed in the medium-rest frame, could be Abraham's or Minkowski's [19,22], or even "arbitrary" [11]; thus leading to the momentum not having a determinate value. Such a result does not make sense physically.

On the other hand, the existing theories are not self-consistent in calculation results. For example, in the dielectric medium Einstein-box thought experiment Barnett concluded in his Letter that the photon in the medium behaves as Abraham's momentum because Minkowski's momentum "would be at odds with Newton's first law of motion" [19], while some others concluded that "the Abraham momentum is not uniquely selected as the momentum of light in this case" [22].

The above indeterminacy in the existing theories can be excluded by imposing physical conditions, just like the EM field solutions are required to satisfy boundary conditions. Such a physical condition is the principle of relativity: The laws of physics are the same in all inertial frames of reference. That means that there is *no preferred inertial frame* for descriptions of physical phenomena. For example, Maxwell equations, global momentum and energy conservation laws, and the principle of Fermat are all valid in any inertial frames, no matter whether the medium is moving or at rest, and no matter whether the space is fully or partially filled with a medium.

One might insist that the medium should define a preferred frame of reference so that there is absolutely no reason why the Fermat's principle is valid in all inertial frames. However in this paper, the principle of relativity is taken as a fundamental hypothesis no matter with or without the existence of a medium.

The momentum of light in a medium can be described by single photon's momentum or by EM momentum. For a plane wave supported by the uniform medium without any dispersion [24,30] and losses, the phase function characterizes the propagation of energy and momentum of light. (1) The light momentum is parallel to the wave vector, and (2) the phase function is a Lorentz invariant (confer Sec. IV). As physical laws, according to the principle of relativity, the above two basic properties are valid in any inertial frames. From this we may expect that the light momentum and energy must constitute a Lorentz covariant four-vector.

The uniform-medium model is the simplest physical model that can be strictly treated mathematically in the Maxwell-equation frame; however, the physical results obtained are fundamental in understanding the physics of light momentum. For example, the Lorentz transformation of photon density in the *isotropic-fluid model* treated by sophisticated field-theory approach [33, Eq. (100)] is exactly the same as that in the uniform-medium model [34, Eq. (44)].

The uniform-medium model has a basic assumption that the dielectric parameters ( $\mu = \mathbf{B}/\mathbf{H}$ and $\varepsilon = \mathbf{D}/\mathbf{E}$ ) are taken to be real scalar constants observed in the medium-rest frame (confer Sec. IV). This assumption is widely used in textbooks of electromagnetism and literature [10,12,19]; however, this assumption never means the uniform medium to be a "rigid body", because all atoms or molecules in dielectric materials used as a uniform medium are always in constant motion or vibration. In fact, the uniform-medium model is strongly supported by the well-known relativity experiment, Fizeau running-water experiment, where the refractive index of the running water in the water-rest frame is a constant.

One might challenge the validity of the argument of "light momentum is parallel to the wave vector", especially when observed in the frames of motion relative to the medium-rest frame, where the moving isotropic medium behaves as being *an*isotropic. In fact, this argument can be easily understood through the property of light propagation and the Lorentz invariance of phase function. Conceptually speaking, the direction of motion of photons is the direction of the light momentum and energy propagation. The phase function defines equi-phase planes of motion (wavefronts), with the wave vector as the normal vector. From one equi-phase plane to another equi-phase plane, the path parallel to the normal vector is the shortest. According to Fermat's principle, light follows the path of least time. Thus the direction of motion of the photons must be parallel to the wave vector, and so must the light momentum. Since the phase function is Lorentz symmetric, this property of light momentum must be valid in all inertial frames. In short, the form of the phase function [confer Eq. (5)] requires the light momentum to be parallel to the wave vector, while the Lorentz invariance of the phase function insures the universal validity of the light-momentum property.

In this paper, by analysis of a plane wave in a moving uniform medium, it is shown that, Minkowski's light momentum is parallel to the wave vector in all inertial frames, and the momentum and energy constitute a Lorentz four-vector, while Abraham's momentum does not have such properties. Thus if the Abraham's momentum is taken as the correct light momentum [19] or an EM momentum postulate [16], an unfavorable physical consequence has to be faced: the global momentum and energy conservation laws, which are fundamental postulates in physics [23], are broken [40]. Fizeau running water experiment is re-analyzed as a support to the Minkowski's momentum.

The photon momentum-energy four-vector is constructed based on invariance of phase and Einstein's light-quantum hypothesis, while the EM momentum-energy four-vector is "classically" constructed based on the covariance of EM fields; the latter, when imposed by Einstein's light-quantum hypothesis, is restored to the former.

In this paper it is also shown for the plane wave that, (*a*) there may be a pseudo-power flow when a medium moves, and the Poynting vector does not necessarily denote the direction of real power flowing; (*b*) the moving medium behaves as a so-called "negative index medium" [41-44] when it moves opposite to the wave vector at a faster-than-dielectric light speed [45]. The physical implication of the puzzling "negative frequency" appearing in the superluminal medium, which results in the question of invariance of phase [45,46], is elucidated.

Some simple examples are given to show: When static electric or magnetic field moves in free-space, there must be pseudo-momentum appearing, and neither Abraham's nor Minkowski's momentum can correctly describe the real EM momentum. As an application of this principle, the classical electron mass-energy paradox is analyzed and resolved.

Finally, in terms of the traditional understanding of the momentum equal to mass multiplied by velocity ($\boldsymbol{p} = m\boldsymbol{v}$) in Newton classical mechanics and the analysis of various simple but basic forms of EM fields, a general EM momentum definition is introduced. From this new definition, it is found for the first time that the traditional "Abraham-type" and "Minkowski-type" momentums can be included in the same momentum formulation for dispersion wave-guiding systems, such as the regular dielectric-filled metallic waveguides, where finite transverse dimensions result in the dispersion of EM momentum, thus leading to "Minkowski-type" momentum appearing at high frequency while "Abraham-type" momentum appearing at relatively low frequency.

The paper is organized as follows. In Sec. II, invariant forms of basic physical quantities for a plane wave in a moving uniform medium are defined based on invariance of phase, such as refractive index, phase velocity, and group velocity. In Sec. III, single photon's momentum is analyzed. In Sec. IV, by use of the plane-wave field solution, EM momentum of light and "negative index medium" effect are analyzed. Finally in Sec V, some conclusions and remarks are given. Resolution of the classical electron EM mass-energy paradox is presented, and a general EM momentum definition is proposed.

## II. REFRACTIVE INDEX, PHASE VELOCITY, AND GROUP VELOCITY

In this section, invariant forms of refractive index, phase velocity, and group velocity are defined based on invariance of phase for a plane wave in a moving uniform medium. An unconventional analysis of the relation between the group velocity and Poynting vector is given.

According to the principle of relativity [47-51], no matter whether the space is partially or fully filled by dielectric materials and no matter what kinds of dielectric properties the filled materials have, all inertial frames are symmetric for descriptions of physical laws. To insure the symmetry the time and space must follow Lorentz transformations which are *independent of* the existence of materials, and so do the EM fields to keep Maxwell equations invariant in form.

Suppose that the frame $X'Y'Z'$ moves with respect to the lab frame $XYZ$ at a constant velocity of $\boldsymbol{\beta}c$, with all corresponding coordinate axes in the same directions and their origins overlapping at $t = t' = 0$, as shown in Fig. 1.

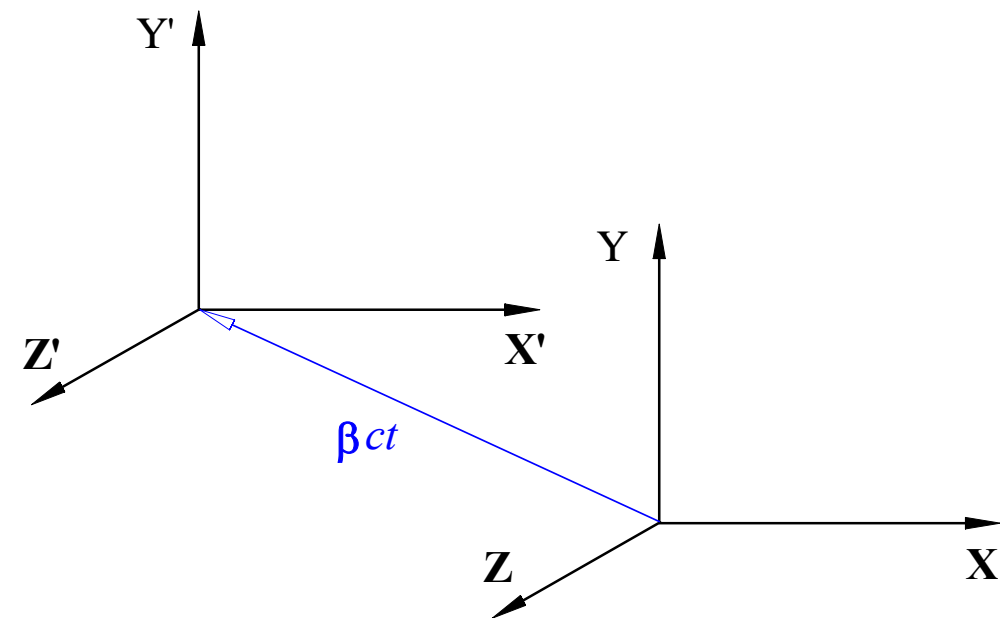

Fig. 1. Two inertial frames of relative motion. $X'Y'Z'$ moves with respect to $XYZ$ at $\boldsymbol{\beta}c$, while $XYZ$ moves with respect to $X'Y'Z'$ at $\boldsymbol{\beta}'c$ (not shown), with $\boldsymbol{\beta}' = -\boldsymbol{\beta}$. Note: $(\gamma\boldsymbol{\beta}, \gamma)$ is the four-vector describing the motion of $X'Y'Z'$, while $(\gamma'\boldsymbol{\beta}', \gamma')$ with $\gamma' = \gamma$ is the four-vector describing the motion of $XYZ$; thus $(\gamma\boldsymbol{\beta}, \gamma)$ and $(\gamma'\boldsymbol{\beta}', \gamma')$ are not the same four-vector, which is an exception in this primed-unprimed symbol usage.

The Lorentz transformation of the time-space four-vector $(\mathbf{x}, ct)$ is given by [47]

$$\mathbf{x} = \mathbf{x}' + \frac{\gamma - 1}{\beta^2}(\boldsymbol{\beta}' \cdot \mathbf{x}')\boldsymbol{\beta}' - \gamma\boldsymbol{\beta}' ct', \tag{1}$$

$$ct = \gamma(ct' - \boldsymbol{\beta}' \cdot \mathbf{x}'). \tag{2}$$

where $c$ is the universal light speed, and $\gamma = (1 - \boldsymbol{\beta}^2)^{-1/2}$ is the time dilation factor. The EM fields **E** and **B**, and **D** and **H** respectively constitute a covariant second-rank anti-symmetric tensor $F^{\alpha\beta}(\mathbf{E}, \mathbf{B})$ and $G^{\alpha\beta}(\mathbf{D}, \mathbf{H})$, of which the Lorentz transformations can be written in intuitive 3D-vector forms, given by [47,48]

$$\begin{pmatrix} \mathbf{E} \\ \mathbf{D} \end{pmatrix} = \gamma \begin{pmatrix} \mathbf{E}' \\ \mathbf{D}' \end{pmatrix} + \gamma\boldsymbol{\beta}' \times \begin{pmatrix} \mathbf{B}'c \\ \mathbf{H}'/c \end{pmatrix} - \frac{\gamma - 1}{\beta^2}\boldsymbol{\beta}' \cdot \begin{pmatrix} \mathbf{E}' \\ \mathbf{D}' \end{pmatrix}\boldsymbol{\beta}', \tag{3-1}$$

with

$$\mathbf{E} \cdot \mathbf{D} = \gamma^2[(\mathbf{E}' \cdot \mathbf{D}') + \beta^2(\mathbf{B}' \cdot \mathbf{H}')]$$

$$-\gamma^2 c\boldsymbol{\beta}' \cdot \left(\mathbf{D}' \times \mathbf{B}' + \frac{\mathbf{E}' \times \mathbf{H}'}{c^2}\right)$$

$$-\gamma^2[(\boldsymbol{\beta}'\cdot\mathbf{B}')(\boldsymbol{\beta}'\cdot\mathbf{H}')+(\boldsymbol{\beta}'\cdot\mathbf{E}')(\boldsymbol{\beta}'\cdot\mathbf{D}')]\,, \tag{3-2}$$

and

$$\begin{pmatrix}\mathbf{B}\\ \mathbf{H}\end{pmatrix}=\gamma\begin{pmatrix}\mathbf{B}'\\ \mathbf{H}'\end{pmatrix}-\gamma\boldsymbol{\beta}'\times\begin{pmatrix}\mathbf{E}'/c\\ \mathbf{D}'c\end{pmatrix}-\frac{\gamma-1}{\beta^2}\boldsymbol{\beta}'\cdot\begin{pmatrix}\mathbf{B}'\\ \mathbf{H}'\end{pmatrix}\boldsymbol{\beta}'\,, \tag{4-1}$$

with

$$\mathbf{B}\cdot\mathbf{H}=\gamma^2[(\mathbf{B}'\cdot\mathbf{H}')+\beta^2(\mathbf{E}'\cdot\mathbf{D}')]$$

$$-\gamma^2 c\boldsymbol{\beta}'\cdot\left(\frac{\mathbf{E}'\times\mathbf{H}'}{c^2}+\mathbf{D}'\times\mathbf{B}'\right)$$

$$-\gamma^2[(\boldsymbol{\beta}'\cdot\mathbf{E}')(\boldsymbol{\beta}'\cdot\mathbf{D}')+(\boldsymbol{\beta}'\cdot\mathbf{B}')(\boldsymbol{\beta}'\cdot\mathbf{H}')]\,. \tag{4-2}$$

The quantities $\mathbf{E}\cdot\mathbf{B}$, $\mathbf{E}^2-(\mathbf{B}c)^2$, $\mathbf{D}\cdot\mathbf{H}$, and $(\mathbf{D}c)^2-\mathbf{H}^2$ are Lorentz invariants [49], and from Eq. (3-2) and Eq. (4-2), we directly know that the energy-density difference between electric and magnetic fields, given by $0.5(\mathbf{E}\cdot\mathbf{D}-\mathbf{B}\cdot\mathbf{H})$, is also a Lorentz invariant [49].

**Refractive index and its Lorentz transformation.** Suppose that there is a plane wave propagating in the medium-rest frame $X'Y'Z'$, and the plane wave has a propagation factor of $\exp(i\Psi')$, where the phase function is given by $\Psi'(\mathbf{x}',t')=\omega't'-n_d'\mathbf{k}'\cdot\mathbf{x}'$, with $\omega'$ $(>0)$ the angular frequency, $n_d'\mathbf{k}'$ the wave vector, $|\mathbf{k}'|=\omega'/c$ required by wave equation, and $n_d'>0$ the refractive index of medium. It is seen from Eqs. (3-1) and (4-1) that the phase function $\Psi(\mathbf{x},t)$ for this plane wave observed in the lab frame *XYZ* must be equal to $\Psi'(\mathbf{x}',t')$ (confer Sec. IV), namely invariance of phase. Thus we have

$$\Psi=\omega t-n_d\mathbf{k}\cdot\mathbf{x}=\omega't'-n_d'\mathbf{k}'\cdot\mathbf{x}'\,, \tag{5}$$

where $n_d\mathbf{k}$ is the wave vector in the lab frame, with $n_d>0$ the refractive index and $|\mathbf{k}|=|\omega/c|$. Note: $\omega$ can be negative [45,46].

From the four-vector covariance of $(\mathbf{x}',ct')$ and the invariance of phase, we conclude that $(n_d'\mathbf{k}',\omega'/c)$ must be a Lorentz four-vector [50]. By setting the time-space four-vector $X^\mu=(\mathbf{x},ct)$ and the wave four-vector $K^\mu=(n_d\mathbf{k},\omega/c)$, Eq. (5) can be written as $(\omega t-n_d\mathbf{k}\cdot\mathbf{x})=g_{\mu\nu}K^\mu X^\nu$ with the metric tensor $g_{\mu\nu}=g^{\mu\nu}=diag(-1,-1,-1,+1)$ [51]. Since $X^\mu$ must be a four-vector, the invariance of phase and the four-vector covariance of $K^\mu$ are equivalent.

From Eqs. (1) and (2) with $n_d'\mathbf{k}'\to\mathbf{x}'$ and $\omega'/c\to ct'$, we obtain $K^\mu=(n_d\mathbf{k},\omega/c)$. Setting $\hat{\mathbf{n}}'=n_d'\mathbf{k}'/|n_d'\mathbf{k}'|$ as the unit wave vector we have

$$\omega=\omega'\gamma(1-n_d'\,\hat{\mathbf{n}}'\cdot\boldsymbol{\beta}')\,,\qquad\text{(Doppler formula)} \tag{6}$$

$$n_d\mathbf{k}=(n_d'\mathbf{k}')+\frac{\gamma-1}{\beta^2}(n_d'\mathbf{k}')\cdot\boldsymbol{\beta}'\boldsymbol{\beta}'-\gamma\boldsymbol{\beta}'\left(\frac{\omega'}{c}\right). \tag{7}$$

From Eq. (6) and the Minkowski-metric-expressed dispersion equation

$$g^{\mu\nu}(K_\mu K_\nu-K_\mu'K_\nu')=0\,, \tag{8-1}$$

namely

$$\left(\frac{\omega}{c}\right)^2-(n_d\mathbf{k})^2=\left(\frac{\omega'}{c}\right)^2-(n_d'\mathbf{k}')^2\,, \tag{8-2}$$

we obtain the refractive index in the lab frame, given by

$$n_d=\frac{\sqrt{(n_d'^2-1)+\gamma^2(1-n_d'\hat{\mathbf{n}}'\cdot\boldsymbol{\beta}')^2}}{|\gamma(1-n_d'\hat{\mathbf{n}}'\cdot\boldsymbol{\beta}')|}\,. \tag{9}$$

From above Eq. (9), we see that the motion of dielectric medium results in an anisotropic refractive index.

It is worthwhile to point out that, the Minkowski-metric dispersion relation Eq. (8) can be easily shown to be equivalent to the Gordon-metric dispersion relation $\Gamma^{\mu\nu}K_\mu K_\nu=0$ [22], where the well-known Gordon metric is given by $\Gamma^{\mu\nu}=g^{\mu\nu}-(1-n_d'^2)B^\mu B^\nu$, with $B^\mu=\partial X^\mu/\partial X'^4=(\gamma\boldsymbol{\beta},\gamma)$ and $B^\mu K_\mu=K_4'=\omega'/c$ (see Attachments-I and II). However the Minkowski-metric dispersion relation is much simpler and more intuitive. [Note: (1) $B^\mu=(\gamma\boldsymbol{\beta},\gamma)$ is the normalized four-velocity of the medium-rest frame $X'Y'Z'$ observed in the lab frame $XYZ$, but when observed in $X'Y'Z'$, $B^\mu=(\gamma\boldsymbol{\beta},\gamma)$ is Lorentz-transformed into $B'^\mu=(\mathbf{0},1)$; confer Fig.1. (2) $B^\mu K_\mu=B'^\mu K_\mu'=\omega'/c$ means the Lorentz invariance of the scalar product $B^\mu K_\mu$, but it never means that $\omega'/c$ is also a Lorentz invariant.]

**Phase velocity and photon's propagation velocity.** It is seen from Eq. (5) that the phase function is symmetric with respect to all inertial frames, independent of which frame the medium is fixed in; accordingly, no frame should make its phase function have any priority in time and space. From this we can conclude: the definitions of equi-phase plane and phase velocity should be symmetric, independent of the choice of inertial frames. Thus the phase velocity can be defined as

$$\boldsymbol{\beta}_{ph}c=\frac{\omega}{|n_d\mathbf{k}|}\hat{\mathbf{n}}=\frac{c}{n_d}\frac{\omega}{|\omega|}\hat{\mathbf{n}}=\beta_{ph}c\,\hat{\mathbf{n}}\,, \tag{10-1}$$

leading to

$$\omega-n_d\mathbf{k}\cdot\boldsymbol{\beta}_{ph}c=0\,, \tag{10-2}$$

where $\hat{\mathbf{n}}=n_d\mathbf{k}/|n_d\mathbf{k}|$ is the unit wave vector in the lab frame, and $\boldsymbol{\beta}_{ph}c$ and $K^\mu$ are related through $K^\mu=(n_d\mathbf{k},\,\omega/c)$ $=\omega(n_d/c)^2(\boldsymbol{\beta}_{ph}c,\,c/n_d^2)$. Note: the definition of the phase velocity $\boldsymbol{\beta}_{ph}c$ is based on the wave four-vector $K^\mu$, while the velocity definition of a *massive particle* is based on the time-space four-vector $X^\mu$. Because the phase velocity $\boldsymbol{\beta}_{ph}c$ is parallel to $n_d\mathbf{k}$, which is a constraint, it cannot be used to construct a "phase velocity four-vector".

At first sight, one might conjecture that $\gamma_{ph}(\boldsymbol{\beta}_{ph}c,\,c)$ could be the "phase velocity four-vector", with $\gamma_{ph}=(1-\beta_{ph}^2)^{-1/2}$; however, on second thoughts one may find that it is not true because $\gamma_{ph}(\boldsymbol{\beta}_{ph}c,\,c)$ and $K^\mu=\omega(n_d/c)^2(\boldsymbol{\beta}_{ph}c,\,c/n_d^2)$ cannot satisfy Lorentz transformations *at the same time*.

From Eq. (5), the equi-phase-plane (wavefront) equation of motion is given by $\omega t-n_d|\mathbf{k}|\hat{\mathbf{n}}\cdot\mathbf{x}=const$, with $\hat{\mathbf{n}}$ as the unit normal vector of the plane, leading to $\omega-n_d|\mathbf{k}|\hat{\mathbf{n}}\cdot(d\mathbf{x}/dt)=0$. Comparing with Eq. (10-1), we obtain $\boldsymbol{\beta}_{ph}c=\hat{\mathbf{n}}\cdot(d\mathbf{x}/dt)\hat{\mathbf{n}}$. Thus we have a physical explanation to $\boldsymbol{\beta}_{ph}c$: the phase velocity is equal to the changing rate of the equi-phase plane's *distance* displacement $\hat{\mathbf{n}}(\hat{\mathbf{n}}\cdot d\mathbf{x})$ over time $dt$, and it is the photon's propagation velocity. Obviously, this photon-velocity definition is consistent with the Fermat's principle in all inertial frames: Light follows the path of least time.

In general, $d\mathbf{x}/dt$ with $|\boldsymbol{\beta}_{ph}c|\le|d\mathbf{x}/dt|<\infty$ in the expression $\boldsymbol{\beta}_{ph}c=\hat{\mathbf{n}}\cdot(d\mathbf{x}/dt)\hat{\mathbf{n}}$ is undetermined unless a definition is given. If $d\mathbf{x}'/dt'=\boldsymbol{\beta}_{ph}'c$ is assigned in the *medium-rest* frame, we call $\mathbf{u}\equiv d\mathbf{x}/dt$ the photon's *apparent* velocity (also known as "ray velocity" in textbooks [52,53]; "apparent" here means "look like but not necessarily real"), with $\gamma_u(\mathbf{u},c)$ its four-velocity. Thus $\boldsymbol{\beta}_{ph}c=(\hat{\mathbf{n}}\cdot\mathbf{u})\hat{\mathbf{n}}$ is form-invariant in all inertial frames.

From the equi-phase-plane equation $\omega t-n_d\mathbf{k}\cdot\mathbf{x}=const$ $\Rightarrow \omega-n_d\mathbf{k}\cdot\mathbf{u}=0$ $\Rightarrow \boldsymbol{\beta}_{ph}c=(\hat{\mathbf{n}}\cdot\mathbf{u})\hat{\mathbf{n}}$, we have introduced the photon's apparent velocity $\mathbf{u}$. The appearance of $\mathbf{u}$ comes from the fact: the photon real velocity is the phase velocity $\boldsymbol{\beta}_{ph}c$, which is defined based on the wave four-vector $K^{\mu}$ instead of the time-space four-vector $X^{\mu}$. From this it follows that, when using the time-space coordinates to describe the motion of a photon, the space coordinates may not reflect the photon's real location, resulting in an illusion. Thus there must be a conversion between the photon's apparent and real locations. This conversion is governed by the photon's real-*vs*-apparent velocity equation $\boldsymbol{\beta}_{ph}c=(\hat{\mathbf{n}}\cdot\mathbf{u})\hat{\mathbf{n}}$. From it we have $\Delta\mathbf{x}_{photon}=(\hat{\mathbf{n}}\cdot\Delta\mathbf{x})\hat{\mathbf{n}}$, where $\Delta\mathbf{x}_{photon}\equiv\boldsymbol{\beta}_{ph}c\Delta t$ is the photon's *real* displacement and $\Delta\mathbf{x}\equiv\mathbf{u}\Delta t$ is its *apparent* displacement, as shown in Fig. 2. Note: $(\Delta\mathbf{x},c\Delta t)$ is a four-vector while $(\Delta\mathbf{x}_{photon},c\Delta t)$ is not, except for in free space, where the "dielectric property" of vacuum medium is symmetric, and the Poynting vector is always parallel to the wave vector in all inertial frames.

From the definitions of $\mathbf{u}$ and $\Delta\mathbf{x}_{photon}$, two conclusions can be drawn. (1) $\mathbf{u}=\mathbf{S}/W_{em}$, with $\mathbf{S}=\mathbf{E}\times\mathbf{H}$ the Poynting vector and $W_{em}=0.5(\mathbf{D}\cdot\mathbf{E}+\mathbf{B}\cdot\mathbf{H})$ the EM energy density [see Eq. (III-24) of Attachment-III]. $\mathbf{S}/W_{em}$ is the so-called "energy velocity" traditionally [53], and it is equal to the phase velocity in the medium-rest frame, but it is *larger than* the phase velocity in general in the lab frame. (2) Photon's Minkowski angular momentum conservation. Photon's momentum is given by $\hbar n_d\mathbf{k}$ [confer Eq. (14)]. Without loss of generality, suppose that the photon is located at $\mathbf{x}=\mathbf{x}'=0$ when $t=t'=0$. Thus we have $\mathbf{x}_{photon}=\Delta\mathbf{x}_{photon}=(\hat{\mathbf{n}}\cdot\Delta\mathbf{x})\hat{\mathbf{n}}=(\hat{\mathbf{n}}\cdot\mathbf{x})\hat{\mathbf{n}}$, and $\mathbf{x}_{photon}\times\hbar n_d\mathbf{k}=0$, namely the photon's angular momentum is conservative in all inertial frames.

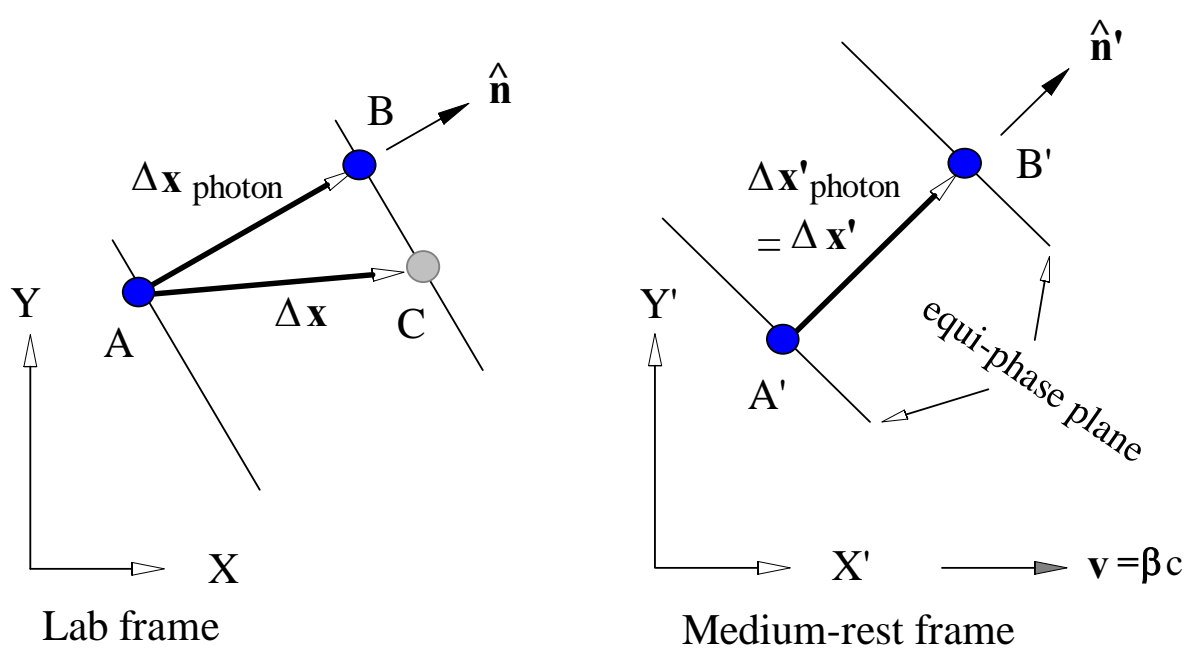

Fig. 2. Photon's real and apparent displacements. The photon propagation velocity is the same as the phase velocity. From Fermat's principle and the principle of relativity, when a photon together with its associated equi-phase plane moves from $A'$ to $B'$ along the unit wave vector $\hat{\mathbf{n}}'$ in the medium rest-frame, it moves from $A$ to $B$ observed in the lab frame. However because the time-space coordinates may not reflect its real location, resulting in an illusion, the photon looks like having moved to $C$ in terms of the time-space Lorentz transformation. Thus the photon's *real* displacement $\Delta\mathbf{x}_{photon}$ only can be converted from its *apparent* displacement $\Delta\mathbf{x}$ through $\Delta\mathbf{x}_{photon}=(\hat{\mathbf{n}}\cdot\Delta\mathbf{x})\hat{\mathbf{n}}$, and the phase velocity is related through $\boldsymbol{\beta}_{ph}c=d\mathbf{x}_{photon}/dt=(\hat{\mathbf{n}}\cdot\mathbf{u})\hat{\mathbf{n}}$, where $\mathbf{u}\equiv d\mathbf{x}/dt$ is the photon's *apparent velocity*, with $\mathbf{u}'=(c/n_d')\hat{\mathbf{n}}'$ in the medium-rest frame. Note: $\boldsymbol{\beta}_{ph}c\,//\,\Delta\mathbf{x}_{photon}$ and $\mathbf{u}\,//\,\Delta\mathbf{x}$; $\Delta\mathbf{x}_{photon}=\Delta\mathbf{x}$ holds if $\mathbf{u}$ and $\boldsymbol{\beta}_{ph}c$ have the same direction.

**Group velocity and its relation with Poynting vector.** The classical definition of group velocity is given by $\mathbf{v}_{gr-c}=\partial\omega/\partial(n_d\mathbf{k})$, defined in the ray-vector direction [48,52]. In this paper we suggest a *modified* definition, given by $\mathbf{v}_{gr}=\hat{\mathbf{n}}\,\partial\omega/\partial|n_d\mathbf{k}|$, defined in the wave-vector direction. Obviously, $\mathbf{v}_{gr}\cdot\hat{\mathbf{n}}=\mathbf{v}_{gr-c}\cdot\hat{\mathbf{n}}$ holds between the two definitions.

From Eq. (5), we know that the form-invariant definition of refractive index $n_d=|n_d\mathbf{k}|/|\omega/c|$ itself also defines a dispersion equation of $(n_d\mathbf{k})^2-(n_d\,\omega/c)^2=0$ for the plane wave. From the modified definition $\mathbf{v}_{gr}\equiv\boldsymbol{\beta}_{gr}c=\hat{\mathbf{n}}\,\partial\omega/\partial|n_d\mathbf{k}|$, we obtain

$$\mathbf{v}_{gr}=\frac{\boldsymbol{\beta}_{ph}c}{1+(\omega/n_d)(\partial n_d/\partial\omega)}. \tag{11}$$

Since the medium is assumed to be non-dispersive in the uniform-medium model, $\partial n_d/\partial\omega=0$ is valid. Thus we have $\mathbf{v}_{gr}=\boldsymbol{\beta}_{ph}c$, namely the group velocity is equal to the phase velocity, parallel to the wave vector.

As we know, for a plane wave in an anisotropic medium the wave vector and Poynting vector usually are not parallel. It has been thought that the group velocity is parallel to the Poynting vector, instead of the wave vector, as shown in the classical electrodynamics textbook by Landau and Lifshitz [52].

The moving isotropic medium becomes an anisotropic medium, as seen in Eq. (9); however, the group velocity we obtained is $\mathbf{v}_{gr}=\boldsymbol{\beta}_{ph}c$, parallel to the wave vector instead of the Poynting vector. Obviously, this is not in agreement with the result in the textbook [52].

Why do we have to modify the group velocity definition? From following analysis, we can see that there is some flaw in the classical definition.

Following the Landau-Lifshitz approach in analysis of a plane wave in an anisotropic lossless medium, with the holding of $(\delta\mathbf{D}\cdot\mathbf{E}-\mathbf{D}\cdot\delta\mathbf{E})+(\delta\mathbf{B}\cdot\mathbf{H}-\mathbf{B}\cdot\delta\mathbf{H})=0$ taken into account for a moving *non-dispersive* uniform medium, from Maxwell equations we obtain $\delta\omega=\mathbf{S}\cdot\delta(n_d\mathbf{k})/W_{em}$, where $\delta(n_d\mathbf{k})$ is an arbitrary infinitesimal change in wave vector, $\mathbf{S}=\mathbf{E}\times\mathbf{H}$ is the Poynting vector, and $W_{em}=0.5(\mathbf{E}\cdot\mathbf{D}+\mathbf{B}\cdot\mathbf{H})$ is the EM energy density. From the mathematical definition of the gradient $\partial\omega/\partial(n_d\mathbf{k})=\mathbf{v}_{gr-c}$, we have $\delta\omega=\mathbf{v}_{gr-c}\cdot\delta(n_d\mathbf{k})$ holding for an arbitrary $\delta(n_d\mathbf{k})$. Comparing $\delta\omega=\mathbf{S}\cdot\delta(n_d\mathbf{k})/W_{em}$ and $\delta\omega=\mathbf{v}_{gr-c}\cdot\delta(n_d\mathbf{k})$, we have $\mathbf{v}_{gr-c}=\mathbf{S}/W_{em}$, namely the classical group velocity is equal to the "energy velocity $\mathbf{S}/W_{em}$" [53], parallel to the Poynting vector. [Note: $\delta\mathbf{D}\cdot\mathbf{E}-\mathbf{D}\cdot\delta\mathbf{E}=0$ and $\delta\mathbf{B}\cdot\mathbf{H}-\mathbf{B}\cdot\delta\mathbf{H}=0$ cannot separately hold for the moving medium because the dielectric tensors are *not* symmetric unless the medium moves along the wave vector.]

However there is a serious flaw for $\mathbf{v}_{gr-c}=\mathbf{S}/W_{em}$, because $|\mathbf{v}_{gr-c}|$ can be greater than the phase velocity $|\mathbf{v}_{ph}|$, which is not physical when considering the fact that all component waves with different frequencies in a signal have the same phase velocity in a non-dispersive lossless medium (confer Sec. IV).

The modified definition $\mathbf{v}_{gr}=\hat{\mathbf{n}}\,\partial\omega/\partial|n_d\mathbf{k}|$, which leads to $\mathbf{v}_{gr}=\boldsymbol{\beta}_{ph}c$ for a non-dispersive medium, has removed the above flaw, which is right the reason why the classical definition of group velocity must be modified.

Since the modified group velocity Eq. (11) is always parallel to the wave vector $n_d\mathbf{k}$ instead of Poynting vector, the Poynting vector does not necessarily denote the direction of power flowing; this is clearly confirmed from the strict EM field solutions given in Sec. IV [confer Eq. (36)].

For a plane wave in an anisotropic medium, the dispersion equation is given by $(n_d\mathbf{k})^2-(n_d\,\omega/c)^2=0$, while the relation between the refractive index $n_d$ and dielectric parameters is described by Fresnel's equation [52], where $n_d$ does not explicitly contain $|n_d\mathbf{k}|$, because $(\partial/\partial|n_d\mathbf{k}|)(n_d\mathbf{k}_i/|n_d\mathbf{k}|)=0$

with $i = x, y, z$ , just like $(\partial/\partial r)(x/r, y/r, z/r) = 0$ in spherical coordinate systems. If there is a dispersion, $n_d$ implicitly contains $\omega$ through dielectric tensors, leading to $\partial n_d / \partial |n_d \mathbf{k}| = (\partial n_d / \partial \omega) v_{gr}$ ; if the medium has no dispersion, $\partial n_d / \partial \omega = 0 \Rightarrow \partial n_d / \partial |n_d \mathbf{k}| = 0$ .

## III. LORENTZ COVARIANCE OF MINKOWSKI'S PHOTON MOMENTUM AND ENERGY

Minkowski and Abraham proposed different formulations of light momentums in a medium, which were both claimed to be supported by experiments [1-3,9,13]. Especially, the Abraham's momentum was claimed to be strongly supported by a recent direct fiber-recoiling observation [13], although various explanations could be given [14,15,17]. For example, the recoiling could be resulting from the transverse radiation force because of an azimuthal asymmetry of refractive index in the fiber [17].

In this section, single photon's momentum in a medium is analyzed based on Einstein light-quantum hypothesis, and it is shown that the Minkowski's photon momentum is strongly supported by four-vector covariance of relativity. Fizeau running water experiment is re-analyzed as a support to the Minkowski's momentum.

For a uniform plane wave, observed in the medium-rest frame the electric field $\mathbf{E}'$ and magnetic field $\mathbf{B}'$ have the relation $\mathbf{B}'c/n'_d = \hat{\mathbf{n}}' \times \mathbf{E}'$ (confer Sec. IV). Thus the Minkowski's and Abraham's EM momentum density vectors can be expressed as

$$\mathbf{g}'_M = \mathbf{D}' \times \mathbf{B}' = \frac{n'_d}{c} (\mathbf{D}' \cdot \mathbf{E}') \hat{\mathbf{n}}' , \qquad (12)$$

$$\mathbf{g}'_A = \frac{\mathbf{E}' \times \mathbf{H}'}{c^2} = \frac{1}{n'_d c} (\mathbf{D}' \cdot \mathbf{E}') \hat{\mathbf{n}}' . \qquad (13)$$

According to Einstein light-quantum hypothesis, the EM energy density $\mathbf{D}' \cdot \mathbf{E}'$ is proportional to the photon's energy $\hbar\omega'$ , while the momentum densities $\mathbf{g}'_M$ and $\mathbf{g}'_A$ are proportional to the photon's momentum. Thus from Eqs. (12) and (13), we obtain the photon's momentum, $p'_M = n'_d \hbar\omega'/c$ for Minkowski's and $p'_A = \hbar\omega'/(n'_d c)$ for Abraham's.

It is interesting to point out that the Minkowski's photon momentum also can be naturally obtained from the covariance of relativity of wave four-vector, as shown below.

Suppose that the Planck constant $\hbar$ is a Lorentz scalar (confer Sec. IV). From the given definition of wave four-vector, $K'^{\mu} = (n'_d \mathbf{k}', \omega'/c)$ multiplied by $\hbar$ , we obtain the photon's momentum-energy four-vector, given by

$$P'^{\mu} = (\hbar n'_d \mathbf{k}', \hbar\omega'/c) = (\mathbf{p}', E'/c) . \qquad (14)$$

In terms of the four-vector structure, $\mathbf{p}'$ must be the momentum; thus we have the photon's momentum in a medium, given by $p' = n'_d E'/c$ , that is right the Minkowski's photon momentum.

From the principle of relativity, we have the invariance of phase, from which we have the covariant wave four-vector. From the wave four-vector combined with Einstein's light-quantum hypothesis, we have the Minkowski's photon momentum, which strongly supports the consistency of Minkowski's momentum expression with the relativity.

In the classical electrodynamics, Fizeau running water experiment is usually taken to be an experimental evidence of the relativistic velocity addition rule [47]. In fact, it also can be taken to be a support to the Minkowski's momentum. To better understand this, let us make a simple analysis, as shown below.

Suppose that the running-water medium is at rest in the $X'Y'Z'$ frame. Since the Minkowski's momentum-energy four-vector $(\hbar n'_d \mathbf{k}', \hbar\omega'/c)$ is covariant, we have Eqs. (8) and (9) holding. Setting $-\boldsymbol{\beta}' = \boldsymbol{\beta} = \beta \hat{\mathbf{n}}'$ (the water moves parallel to the wave vector), from Eq. (9) we have the refractive index in the lab frame, given by

$$n_d = \left| \frac{n'_d + \beta}{1 + n'_d \beta} \right| . \qquad (15)$$

Thus from Eq. (10-1), the light speed in the running water, observed in the lab frame, is given by

$$|\boldsymbol{\beta}_{ph} c| = \frac{c}{n_d} \approx \frac{c}{n'_d} \left[ 1 + \beta \left( n'_d - \frac{1}{n'_d} \right) \right], \quad \text{for } |\beta| << 1 \qquad (16)$$

which is the very formula confirmed by Fizeau experiment (confer Attachment-III). When the water runs along (opposite to) the wave vector direction, we have $\beta > 0$ ($\beta < 0$) and the light speed is increased (reduced).

The combination of Newton's first law with Einstein's mass-energy equivalence [10-12] is an often-used argument to support Abraham's photon momentum. But Abraham's photon momentum, smaller in dielectric than in vacuum, cannot be obtained from the wave four-vector in a covariant manner, and it is not consistent with the relativity.

In the total-momentum model [19, Eq. (7)], Abraham's and Minkowski's momentums are, respectively, a component of the same total momentum. When applying this model to analysis of the Einstein-box thought experiment [19], the total momentum is reduced into Abraham's momentum observed in the medium-rest frame. Thus the compatibility of this total-momentum model with the special relativity is also called into question [40].

One might use different ways to define a photon's energy in a medium. In the medium-rest frame, the dispersion equation, directly resulting from second-order wave equation (instead of the invariance of phase), is given by [48]

$$(n'_d \omega'/c)^2 - (n'_d \mathbf{k}')^2 = 0 , \qquad (17)$$

which actually is the definition of refractive index. This dispersion relation is thought to be the characterization of the relation between EM energy and momentum, and the photon's energy in a medium is supposed to be $n'_d \hbar\omega'$ to keep a zero rest energy [48, Sec. 3.1*a*]. However it should be noted that, although Eq. (17) is Lorentz invariant in form, $(n'_d \mathbf{k}', n'_d \omega'/c)$ is not a Lorentz covariant four-vector, since only $K'^{\mu} = (n'_d \mathbf{k}', \omega'/c)$ is; except for $n'_d = 1$ . If using $(n'_d \hbar\mathbf{k}', n'_d \hbar\omega'/c)$ to define the photon's momentum-energy four-vector, then it is not Lorentz covariant.

Thus it is justifiable to define $\hbar K'^{\mu} = (n'_d \hbar\mathbf{k}', \hbar\omega'/c)$ as the photon's momentum-energy four-vector, as done in Eq. (14), because $\hbar K'^{\mu}$ is covariant, with Eq. (17) as a natural result.

## IV. NOVEL PROPERTIES OF A PLANE WAVE IN A MOVING UNIFORM MEDIUM

In this section, we will show some novel properties, including the four-vector Lorentz covariance of Minkowski's EM momentum and energy, for a plane wave in a moving lossless, non-conducting, non-dispersive, isotropic uniform medium, which is the simplest strict solution to Maxwell equations [48].

Suppose that the plane-wave solution in the medium-rest frame $X'Y'Z'$ is given by

$$(\mathbf{E}', \mathbf{B}', \mathbf{D}', \mathbf{H}') = (\mathbf{E}'_0, \mathbf{B}'_0, \mathbf{D}'_0, \mathbf{H}'_0) \exp(i\Psi') , \qquad (18)$$

where $\Psi' = (\omega' t' - n'_d \mathbf{k}' \cdot \mathbf{x}')$, with $\omega' > 0$, and $(\mathbf{E}'_0, \mathbf{B}'_0, \mathbf{D}'_0, \mathbf{H}'_0)$ are real constant amplitude vectors. $\mathbf{D}' = \varepsilon' \mathbf{E}'$ and $\mathbf{B}' = \mu' \mathbf{H}'$ hold, where $\varepsilon' > 0$ and $\mu' > 0$ are the constant dielectric permittivity and permeability, with $\varepsilon' \mu' = n'^2_d / c^2$. Thus $(\mathbf{E}', \mathbf{B}', \hat{\mathbf{n}}')$ and $(\mathbf{D}', \mathbf{H}', \hat{\mathbf{n}}')$ are, respectively, two sets of right-hand orthogonal vectors, with $\mathbf{E}' = (c/n'_d)\mathbf{B}' \times \hat{\mathbf{n}}'$ and $\mathbf{H}' = \hat{\mathbf{n}}' \times (c/n'_d)\mathbf{D}'$, resulting from Maxwell equations.

Inserting Eq. (18) into Eqs. (3-1) and (4-1), we obtain the plane-wave solution in the lab frame $XYZ$, given by

$$(\mathbf{E}, \mathbf{B}, \mathbf{D}, \mathbf{H}) = (\mathbf{E}_0, \mathbf{B}_0, \mathbf{D}_0, \mathbf{H}_0)\exp(i\Psi), \tag{19}$$

where $\exp(i\Psi) = \exp(i\Psi')$ must hold for any time-space points, and $(\mathbf{E}_0, \mathbf{B}_0, \mathbf{D}_0, \mathbf{H}_0)$ are given by

$$\begin{pmatrix} \mathbf{E}_0 \\ \mathbf{H}_0 \end{pmatrix} = \gamma(1 - n'_d \hat{\mathbf{n}}' \cdot \boldsymbol{\beta}') \begin{pmatrix} \mathbf{E}'_0 \\ \mathbf{H}'_0 \end{pmatrix} + \left( \gamma n'_d \hat{\mathbf{n}}' - \frac{\gamma - 1}{\beta^2} \boldsymbol{\beta}' \right) \begin{pmatrix} \boldsymbol{\beta}' \cdot \mathbf{E}'_0 \\ \boldsymbol{\beta}' \cdot \mathbf{H}'_0 \end{pmatrix}, \tag{20}$$

$$\begin{pmatrix} \mathbf{B}_0 \\ \mathbf{D}_0 \end{pmatrix} = \gamma\left(1 - \frac{1}{n'_d} \hat{\mathbf{n}}' \cdot \boldsymbol{\beta}'\right) \begin{pmatrix} \mathbf{B}'_0 \\ \mathbf{D}'_0 \end{pmatrix} + \left( \gamma \frac{1}{n'_d} \hat{\mathbf{n}}' - \frac{\gamma - 1}{\beta^2} \boldsymbol{\beta}' \right) \begin{pmatrix} \boldsymbol{\beta}' \cdot \mathbf{B}'_0 \\ \boldsymbol{\beta}' \cdot \mathbf{D}'_0 \end{pmatrix}. \tag{21}$$

Note: All quantities are real in Eqs. (20) and (21) and the transformations are "synchronous"; for example, $\mathbf{E}_0$ is expressed only in terms of $\mathbf{E}'_0$. All field quantities have the same phase factor, no matter in the medium-rest frame or lab frame. It is clearly seen from Eqs. (18)-(21), the invariance of phase, $\Psi = \Psi'$, is a natural result.

In the following analysis, formulas are derived in the lab frame, because they are invariant in forms in all inertial frames. Although the fields given by Eqs. (18) and (19) are complex, all equations and formulas obtained are checked with complex and real fields and they are valid.

Under Lorentz transformations, the Maxwell equations keep the same forms as in the medium-rest frame, given by

$$\nabla \times \mathbf{E} = -\partial \mathbf{B}/\partial t, \qquad \nabla \cdot \mathbf{D} = \rho, \tag{22}$$

$$\nabla \times \mathbf{H} = \mathbf{J} + \partial \mathbf{D}/\partial t, \qquad \nabla \cdot \mathbf{B} = 0, \tag{23}$$

with $\mathbf{J} = 0$ and $\rho = 0$ for the plane wave. From above, we have

$$\omega \mathbf{B} = n_d \mathbf{k} \times \mathbf{E}, \quad \text{and} \quad \omega \mathbf{D} = -n_d \mathbf{k} \times \mathbf{H}, \tag{24}$$

leading to $\mathbf{D} \cdot \mathbf{E} = \mathbf{B} \cdot \mathbf{H}$, namely the electric energy density is equal to the magnetic energy density, which is valid in all inertial frames.

From Eqs (20) and (21), we can directly obtain some intuitive expressions for examining the anisotropy of space relations of EM fields observed in the lab frame, given by

$$\mathbf{D} \cdot \hat{\mathbf{n}} = \mathbf{D}' \cdot \hat{\mathbf{n}}' = 0, \qquad \mathbf{B} \cdot \hat{\mathbf{n}} = \mathbf{B}' \cdot \hat{\mathbf{n}}' = 0, \tag{25}$$

$$\mathbf{E} \cdot \hat{\mathbf{n}} = \frac{\gamma(n'^2_d - 1)(\mathbf{E}' \cdot \boldsymbol{\beta}')}{\sqrt{(n'^2_d - 1) + \gamma^2 (1 - n'_d \hat{\mathbf{n}}' \cdot \boldsymbol{\beta}')^2}}, \tag{26}$$

$$\mathbf{H} \cdot \hat{\mathbf{n}} = \frac{\gamma(n'^2_d - 1)(\mathbf{H}' \cdot \boldsymbol{\beta}')}{\sqrt{(n'^2_d - 1) + \gamma^2 (1 - n'_d \hat{\mathbf{n}}' \cdot \boldsymbol{\beta}')^2}}, \tag{27}$$

and

$$\mathbf{E} \cdot \mathbf{B} = \mathbf{E}' \cdot \mathbf{B}' = 0, \qquad \mathbf{D} \cdot \mathbf{H} = \mathbf{D}' \cdot \mathbf{H}' = 0, \tag{28}$$

$$\mathbf{E} \cdot \mathbf{H} = \gamma^2 (n'^2_d - 1)(\boldsymbol{\beta}' \cdot \mathbf{E}')(\boldsymbol{\beta}' \cdot \mathbf{H}'), \tag{29}$$

$$\mathbf{D} \cdot \mathbf{B} = -\gamma^2 \left(1 - \frac{1}{n'^2_d}\right)(\boldsymbol{\beta}' \cdot \mathbf{D}')(\boldsymbol{\beta}' \cdot \mathbf{B}'), \tag{30}$$

$$\mathbf{D} \cdot \mathbf{E} = \gamma^2 (1 - n'_d \hat{\mathbf{n}}' \cdot \boldsymbol{\beta}') \left(1 - \frac{1}{n'_d} \hat{\mathbf{n}}' \cdot \boldsymbol{\beta}'\right) \mathbf{D}' \cdot \mathbf{E}', \tag{31}$$

$$\mathbf{B} \cdot \mathbf{H} = \gamma^2 (1 - n'_d \hat{\mathbf{n}}' \cdot \boldsymbol{\beta}') \left(1 - \frac{1}{n'_d} \hat{\mathbf{n}}' \cdot \boldsymbol{\beta}'\right) \mathbf{B}' \cdot \mathbf{H}', \tag{32}$$

and

$$\mathbf{E} = (\hat{\mathbf{n}} \cdot \mathbf{E})\hat{\mathbf{n}} - \beta_{ph} c\, \hat{\mathbf{n}} \times \mathbf{B}, \tag{33}$$

$$\mathbf{H} = (\hat{\mathbf{n}} \cdot \mathbf{H})\hat{\mathbf{n}} + \beta_{ph} c\, \hat{\mathbf{n}} \times \mathbf{D}. \tag{34}$$

It is seen from above that $\mathbf{E} \perp \mathbf{B}$, $\mathbf{B} \perp \hat{\mathbf{n}}$, $\mathbf{D} \perp \mathbf{H}$ and $\mathbf{D} \perp \hat{\mathbf{n}}$ hold in the lab frame, but $\mathbf{E} /\!/ \mathbf{D}$, $\mathbf{B} /\!/ \mathbf{H}$, $\mathbf{E} \perp \mathbf{H}$, $\mathbf{D} \perp \mathbf{B}$, $\mathbf{E} \perp \hat{\mathbf{n}}$, and $\mathbf{H} \perp \hat{\mathbf{n}}$ usually do not hold any more.

**Pseudo-power flow caused by a moving medium.** From Eq. (24), we obtain Minkowski's EM momentum and Poynting vector, given by

$$\mathbf{D} \times \mathbf{B} = \left(\frac{\mathbf{D} \cdot \mathbf{E}}{\omega}\right) n_d \mathbf{k} = \left(\frac{n_d}{c}\right)^2 (\mathbf{D} \cdot \mathbf{E})\, \mathbf{v}_{gr}, \tag{35}$$

$$\mathbf{E} \times \mathbf{H} = v_{gr} \, [\, v_{gr} (\mathbf{D} \times \mathbf{B}) - (\hat{\mathbf{n}} \cdot \mathbf{H})\mathbf{B} - (\hat{\mathbf{n}} \cdot \mathbf{E})\, \mathbf{D} \,], \tag{36}$$

where $\mathbf{v}_{gr}$, with $v_{gr} = \mathbf{v}_{gr} \cdot \hat{\mathbf{n}}$ and $|v_{gr}| = c/n_d$, is the group velocity obtained from Eq. (11) with $\partial n_d / \partial \omega = 0$, which is equal to the phase velocity $\boldsymbol{\beta}_{ph} c$, as defined by Eq. (10-1). Note: The Minkowski's momentum $\mathbf{D} \times \mathbf{B}$ has the same direction as the wave vector $n_d \mathbf{k}$, while the Poynting vector $\mathbf{E} \times \mathbf{H}$ has three components: one in $\mathbf{D} \times \mathbf{B}$-direction, one in $\mathbf{B}$-direction, and one in $\mathbf{D}$-direction; the latter two are perpendicular to the group velocity $\mathbf{v}_{gr}$.

We can divide the Poynting vector $\mathbf{S} = \mathbf{E} \times \mathbf{H}$ into two parts, $\mathbf{S} = \mathbf{S}_{power} + \mathbf{S}_{pseu}$, which are perpendicular each other, given by

$$\mathbf{S}_{power} = v^2_{gr} \, (\mathbf{D} \times \mathbf{B}) = (\mathbf{D} \cdot \mathbf{E})\, \mathbf{v}_{gr} = W_{em} \mathbf{v}_{gr}, \tag{37}$$

$$\mathbf{S}_{pseu} = -v_{gr} \, [\, (\hat{\mathbf{n}} \cdot \mathbf{H})\mathbf{B} + (\hat{\mathbf{n}} \cdot \mathbf{E})\, \mathbf{D} \,], \tag{38}$$

where $\mathbf{S}_{power}$ is associated with the total EM energy density $W_{em} = 0.5(\mathbf{D} \cdot \mathbf{E} + \mathbf{B} \cdot \mathbf{H})$ and the group velocity $\mathbf{v}_{gr}$.

$\mathbf{S}_{power}$ carries the total EM energy moving at $\mathbf{v}_{gr}$, and it is a real power flow. $\mathbf{S}_{pseu}$ is perpendicular to the group velocity, and it is a pseudo-power flow. According to Eq. (37), the energy velocity is supposed to be equal to the group velocity.

The essential difference in physics between $\mathbf{S}_{power}$ and $\mathbf{S}_{pseu}$ can be seen from the divergence theorem. The divergence of $\mathbf{S}_{power}$ is given by

$$\nabla \cdot \mathbf{S}_{power} = \nabla \cdot [(\mathbf{D} \cdot \mathbf{E})\, \boldsymbol{\beta}_{gr} c] = (\mathbf{D}_0 \cdot \mathbf{E}_0)[\nabla \cos^2(\omega t \text{-} n_d \mathbf{k} \cdot \mathbf{x})] \cdot \mathbf{v}_{gr} \neq 0, \tag{39}$$

which means that there is a power flowing in and out in the differential box, but the time average $<\nabla\cdot\mathbf{S}_{power}>=0$, meaning that the powers going in and out are the same in average, with no net energy left in the box. In contrast, $\nabla\cdot\mathbf{S}_{pseu}\equiv 0$ holds for any time resulting from $\nabla\cdot\mathbf{B}=0$, $\nabla\cdot\mathbf{D}=0$, $\mathbf{B}\perp\mathbf{k}$, and $\mathbf{D}\perp\mathbf{k}$, which means that there is no power flowing at any time for any places.

Since $\mathbf{S}_{power}$ and $\mathbf{S}_{pseu}$ are mutually perpendicular, $\left|\mathbf{S}/W_{em}\right|>\left|\mathbf{S}_{power}/W_{em}\right|=\left|\boldsymbol{\beta}_{ph}c\right|$ holds for $\mathbf{S}_{pseu}\neq 0$. If $\mathbf{S}/W_{em}$ were defined as the group velocity or energy velocity [52,53], then the group velocity or energy velocity would be greater than the phase velocity, which is not physical because all component waves with different frequencies for a signal have the same phase velocity in a non-dispersive lossless medium.

According to Eqs. (26) and (27), the pseudo-power flow $\mathbf{S}_{pseu}$ vanishes in free space, or when the dielectric medium moves parallel to the wave vector $\hat{\mathbf{n}}'$.

It is seen from above analysis that Poynting vector does not necessarily denote a real power flowing. However such a phenomenon seems neglected in the physics community, in view of the fact that the Abraham's momentum, defined through the Poynting vector, is taken as an EM momentum postulate, as proposed by Mansuripur and Zakharian [16].

In summary, we can make some conclusions for the power flowing. (1) Observed in any inertial frames, the Minkowski's EM momentum $\mathbf{D}\times\mathbf{B}$ is parallel to the wave vector [see Eq. (35)], which is completely in agreement with the Fermat's principle and the principle of relativity, as indicated in the Introduction section. (2) When a medium moves, the Poynting vector $\mathbf{E}\times\mathbf{H}$ consists of two parts: one is parallel to the group velocity, and it is a real power flow; the other is perpendicular to the group velocity, and it is a pseudo-power flow [see Eq. (36)].

**Lorentz covariance of Minkowski's EM momentum and energy, and invariance of Planck constant.** We have shown the Lorentz covariance of Minkowski's photon momentum and energy from the wave four-vector combined with Einstein's light-quantum hypothesis. This covariance suggests us that there should be a covariant EM momentum-energy four-vector, given by

$$\bar{P}^{\mu}=(\bar{\mathbf{p}}_{em},\bar{E}_{em}/c), \qquad (40)$$

where $\bar{\mathbf{p}}_{em}$ and $\bar{E}_{em}$ are, respectively, the EM momentum and energy for a single "EM-field cell" or "photon", given by

$$\bar{\mathbf{p}}_{em}=\frac{\mathbf{D}\times\mathbf{B}}{N_p}, \qquad \bar{E}_{em}=\frac{\mathbf{D}\cdot\mathbf{E}}{N_p}, \qquad (41)$$

with $N_p$ the "EM-field-cell number density" or "photon number density" in volume. The EM momentum-energy four-vector and wave four-vector are related through $\bar{P}^{\mu}=(\bar{E}_{em}/\omega)K^{\mu}$, with $\bar{E}_{em}/\omega$ corresponding to $\hbar$ physically.

The four-vector $\bar{P}^{\mu}=(\bar{\mathbf{p}}_{em},\bar{E}_{em}/c)$ is required to follow a four-vector Lorentz transformation given by Eqs. (1) and (2), while the EM fields must follow a second-rank tensor Lorentz transformation given by Eqs. (3-1) and (4-1), or Eqs. (20) and (21) for a plane wave. From the four-vector Lorentz transformation of $\bar{P}^{\mu}=(\bar{E}_{em}/\omega)K^{\mu}$, with the invariance of phase $\Psi=\Psi'$ considered, we have

$$\frac{N_p\omega}{N_p'\omega'}=\frac{\mathbf{D}_0\cdot\mathbf{E}_0}{\mathbf{D}_0'\cdot\mathbf{E}_0'}. \qquad (42)$$

The above equation has a clear physical explanation that the Doppler factor of EM energy density is equal to the product of the Doppler factors of EM-field-cell density and frequency.

From the second-rank tensor Lorentz transformation of $\mathbf{D}_0$ and $\mathbf{E}_0$ given by Eqs. (20) and (21), we obtain the transformation of EM energy density, given by

$$\frac{\mathbf{D}_0\cdot\mathbf{E}_0}{\mathbf{D}_0'\cdot\mathbf{E}_0'}=\left[\gamma(1-n_d'\hat{\mathbf{n}}'\cdot\boldsymbol{\beta}')\right]\left[\gamma\left(1-\frac{1}{n_d'}\hat{\mathbf{n}}'\cdot\boldsymbol{\beta}'\right)\right], \qquad (43)$$

namely Eq. (31). Comparing with Eq. (6), we know that $\gamma(1-n_d'\hat{\mathbf{n}}'\cdot\boldsymbol{\beta}')$ is the frequency Doppler factor. In free space, the above equation is reduced to Einstein's result [50]: $\left|\mathbf{E}\right|=\gamma(1-\hat{\mathbf{n}}'\cdot\boldsymbol{\beta}')\left|\mathbf{E}'\right|$.

Inserting Eq. (43) into Eq. (42), we obtain the transformation of EM-field-cell density, given by

$$N_p=\gamma\left(1-\frac{1}{n_d'}\hat{\mathbf{n}}'\cdot\boldsymbol{\beta}'\right)N_p'. \qquad (44\text{-}1)$$

So far we have finished the proof of the covariance of $(\bar{\mathbf{p}}_{em},\bar{E}_{em}/c)$ by resorting to a parameter of $N_p$, so-called "EM-field-cell density". Actually, we do not have to know what the specific value of $N_p'$ or $N_p$ is, but the ratio of $N_p/N_p'$, and $\bar{P}^{\mu}=(\bar{\mathbf{p}}_{em},\bar{E}_{em}/c)$ is pure "classical", without Planck constant $\hbar$ involved. However, $N_p$ must be the "photon density" when Einstein's light-quantum hypothesis is imposed.

Mathematically speaking, the existence of the covariance of $\bar{P}^{\mu}=(\bar{\mathbf{p}}_{em},\bar{E}_{em}/c)$ is apparent. $\bar{P}^{\mu}$ and $K^{\mu}$ are "parallel", just different by a factor of $(\bar{E}_{em}/\omega)$ which contains an introduced parameter $N_p$ to make the transformation hold.

It is seen from Eqs. (41) and (42) that $(\bar{E}_{em}/\omega)=(\mathbf{D}\cdot\mathbf{E})/(N_p\omega)$ is a Lorentz invariant, and $(\bar{E}_{em}/\omega)=\hbar$ holds when Einstein's light-quantum hypothesis is imposed. Thus the Planck constant $\hbar$ must be a Lorentz invariant. In other words, Einstein's light-quantum hypothesis requires the Lorentz invariance of Planck constant for a plane wave. Therefore, the construction of photon's momentum-energy four-vector, Eq. (14) in Sec. III, is well grounded.

If a volume $dV_{light}'$ in the medium-rest frame moves along the wave vector $n_d'\mathbf{k}'$ at the light speed $(c/n_d')$, then there are no photons that cross its boundary, and the photon number within $dV_{light}'$ keeps constant. In such a case, as shown in Fig. 3, the transformation of the moving volume, termed *light volume*, is given by

$$\frac{dV_{light}'}{dV_{light}}=\gamma\left(1-\frac{1}{n_d'}\hat{\mathbf{n}}'\cdot\boldsymbol{\beta}'\right). \qquad (44\text{-}2)$$

Compared with Eq. (44-1), we find that

$$N_p dV_{light}=N_p'\,dV_{light}' \qquad (44\text{-}3)$$

is a Lorentz invariant, namely the photon number in the light volume is Lorentz invariant. Thus we have the total momentum-energy four-vector in the light volume, given by

$$\bar{P}^{\mu}(N_p dV_{light})=(\mathbf{D}\times\mathbf{B},\mathbf{D}\cdot\mathbf{E}/c)dV_{light}, \qquad (44\text{-}4)$$

or

$$\int_{V_{light}:\ t=const}(\bar{P}^{\mu}N_p)dV=\int_{V_{light}:\ t=const}(\mathbf{D}\times\mathbf{B},\mathbf{D}\cdot\mathbf{E}/c)dV. \qquad (44\text{-}5)$$

The invariance of $N_p dV_{light}$ implies that observed in any inertial frames, all the $N_p dV_{light}$ photons are frozen inside the light volume $dV_{light}$. This result is completely in agreement with the argument of "light momentum is parallel to the wave vector" in all inertial frames, as stated in the Introduction section.

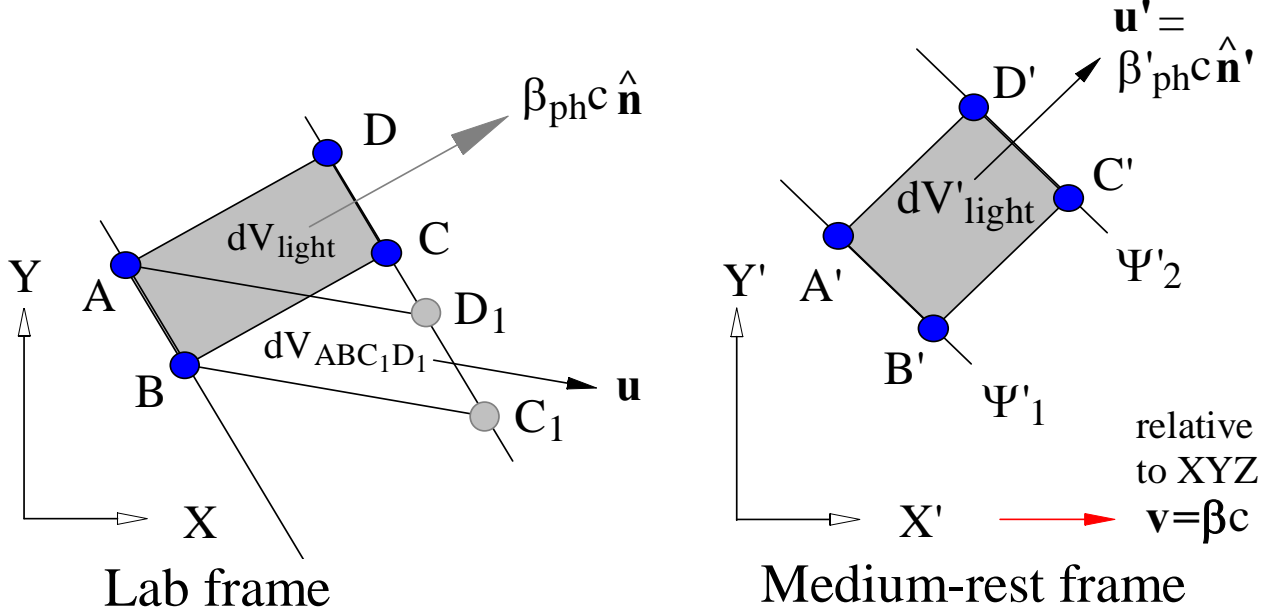

Fig. 3. Transformation of light volumes. From the phase function $\Psi = \omega t - n_d \mathbf{k}\cdot\mathbf{x}$, we have $\boldsymbol{\beta}_{ph}c = (\hat{\mathbf{n}}\cdot\mathbf{u})\hat{\mathbf{n}}$, where the phase velocity $\boldsymbol{\beta}_{ph}c$ is also the photon's *real* velocity and $\mathbf{u}$ is the photon's *apparent* velocity, with $\gamma_u(\mathbf{u},c)$ a four-vector. The Lorentz transformation of light volumes must be performed with the help of $\gamma_u(\mathbf{u},c)$. Suppose that a plane wave propagates along the unit wave vector $\hat{\mathbf{n}}'$ in the medium-rest frame $X'Y'Z'$, which moves at $\mathbf{v}=\boldsymbol{\beta}c$ relatively to the lab frame $XYZ$. On the equi-phase planes $\Psi_1'$ and $\Psi_2'$ $(\neq \Psi_1')$, observed at the same time, there are four photons $A'$, $B'$, $C'$, and $D'$, with $A'D'$ and $B'C'$ parallel to $\hat{\mathbf{n}}'$; thus the rectangle $A'B'C'D'$ can be used to denote the light volume $dV'_{light}$ (upright hexahedron), moving at $\mathbf{u}'=\boldsymbol{\beta}'_{ph}c$. It is assumed that photon $A'$ ($A$) is located at $\mathbf{x}=\mathbf{x}'=0$ when $t=t'=0$. Observed in the lab frame $XYZ$, in terms of the time-space Lorentz transformation, $D_1$ instead of $D$ corresponds to $D'$, and $C_1$ instead of $C$ corresponds to $C'$ although they remain on the same equi-phase plane, resulting in an *apparent* light volume $dV_{ABC_1D_1}$, moving at $\mathbf{u}$. However the principle of relativity requires that the real light volume must be the upright hexahedron $dV_{light}$, instead of the parallel-hexahedron $dV_{ABC_1D_1}$. $dV_{light}$ and $dV_{ABC_1D_1}$ have the same bottom area and height, and their volumes are equal, namely $dV_{light} = dV_{ABC_1D_1}$. Considering that $\gamma_{u'}dV'_{light} = \gamma_u dV_{ABC_1D_1}$ is a Lorentz invariant, we have $dV'_{light}/dV_{light} = \gamma_u/\gamma_{u'}$. $\gamma_u(\mathbf{u},c)$ in $XYZ$ and $\gamma_{u'}(\mathbf{u}',c)$ in $X'Y'Z'$ are the same four-velocity, and from Lorentz transformation we have $\gamma_u c = \gamma(\gamma_{u'}c - \boldsymbol{\beta}'\cdot\gamma_{u'}\mathbf{u}')$ with $\boldsymbol{\beta}'=-\boldsymbol{\beta}$, $\Rightarrow$ $\gamma_u/\gamma_{u'} = \gamma(1-\boldsymbol{\beta}'\cdot\mathbf{u}'/c)$, $\Rightarrow$ $\gamma_u/\gamma_{u'} = \gamma(1-\boldsymbol{\beta}'\cdot\hat{\mathbf{n}}'/n_d')$. From this, we have $dV'_{light}/dV_{light} = \gamma(1-\boldsymbol{\beta}'\cdot\hat{\mathbf{n}}'/n_d')$, namely Eq. (44-2). (Note that the direction of $\mathbf{u}$ is exaggerated to show the difference between $dV_{light}$ and $dV_{ABC_1D_1}$.)

Inserting Eq. (44-2) and Eq. (6) into Eq. (43), we have

$$\frac{(\mathbf{D}\cdot\mathbf{E})dV_{light}}{\omega} = \frac{(\mathbf{D}'\cdot\mathbf{E}')dV'_{light}}{\omega'}, \tag{44-6}$$

which is also a Lorentz invariant; namely the energy in a light volume and the frequency transform in the same law. This result, which is obtained in the moving medium, is exactly the same as that obtained by Einstein in *free-space* [50].

From above analysis, we can draw the following conclusions:

(1) Minkowski's momentum per unit EM-field cell, $N_p^{-1}(\mathbf{D}\times\mathbf{B})$, is Lorentz covariant, as the space component of the momentum-energy four-vector $P^\mu = N_p^{-1}(\mathbf{D}\times\mathbf{B},\mathbf{D}\cdot\mathbf{E}/c)$, just like the Minkowski's photon momentum $\hbar n_d\mathbf{k}$ is Lorentz covariant, as the space component of the momentum-energy four-vector $P^\mu = (\hbar n_d\mathbf{k},\hbar\omega/c)$. When Einstein's light-quantum hypothesis $N_p^{-1}\mathbf{D}\cdot\mathbf{E}=\hbar\omega$ is imposed in the former, the two four-vectors become the same, namely $N_p^{-1}(\mathbf{D}\times\mathbf{B},\mathbf{D}\cdot\mathbf{E}/c)$ $=(\hbar n_d\mathbf{k},\hbar\omega/c)$.

(2) There are two forms of momentum-energy four-vectors: (*a*) the momentum and energy in a single EM-field cell or photon constitute a four-vector, namely $N_p^{-1}(\mathbf{D}\times\mathbf{B},\mathbf{D}\cdot\mathbf{E}/c)$ or $(\hbar n_d\mathbf{k},\hbar\omega/c)$ is a four-vector, and (*b*) the total momentum and energy in a light volume constitute a four-vector, namely $(\mathbf{D}\times\mathbf{B},\mathbf{D}\cdot\mathbf{E}/c)dV_{light}$ is also a four-vector. However, the momentum and energy densities themselves cannot directly constitute a four-vector, namely $(\mathbf{D}\times\mathbf{B},\mathbf{D}\cdot\mathbf{E}/c)$ or $N_p(\hbar n_d\mathbf{k},\hbar\omega/c)$ is never a Lorentz four-vector.

(3) Planck constant is a Lorentz invariant, which is a strict relativistic result for the plane wave in a moving uniform medium.

**Indeterminacy of EM momentum conservation equations, and the issue of momentum transfer.** As we know, in the medium-rest frame the isotropic linear relations $\mathbf{D}'=\varepsilon'\mathbf{E}'$ and $\mathbf{B}'=\mu'\mathbf{H}'$ hold. However such relations do not hold in the lab frame because of the anisotropy of a moving medium. Thus the stress tensor and momentum conservation equation in the lab frame only can be derived based on the invariance of Maxwell equations.

From Eqs. (22) and (23), with $\mathbf{J}=0$ and $\rho=0$ taken into account we have [47]

$$\frac{\partial(\mathbf{D}\times\mathbf{B})}{\partial t} = (\nabla\cdot\mathbf{D})\mathbf{E} - \mathbf{D}\times(\nabla\times\mathbf{E})$$
$$+(\nabla\cdot\mathbf{B})\mathbf{H} - \mathbf{B}\times(\nabla\times\mathbf{H}). \tag{45}$$

Since $\nabla\cdot\mathbf{D}=0$, $\nabla\cdot\mathbf{B}=0$, $\mathbf{B}\perp\hat{\mathbf{n}}$, and $\mathbf{D}\perp\hat{\mathbf{n}}$ hold, we have

$$\frac{\partial(\mathbf{D}\times\mathbf{B})}{\partial t} = -\nabla\cdot\overleftrightarrow{\mathbf{T}}_M, \tag{46}$$

where the symmetric Minkowski's stress tensor is given by

$$\overleftrightarrow{\mathbf{T}}_M = \overleftrightarrow{\mathbf{I}}(\mathbf{D}\cdot\mathbf{E}), \tag{47}$$

with $\overleftrightarrow{\mathbf{I}}$ the unit tensor. Note: Eq. (46) is valid for both real and complex fields although only the real fields are physical.

Eq. (46) is the Minkowski's momentum conservation equation for a plane wave in a moving medium, and it is invariant in form in all inertial frames together with the Maxwell equations.

Now let us take a look of the physical implication of the stress tensor. Consider the EM momentum in a given dielectric volume. From Eq. (46), using divergence theorems for a tensor and a vector, with $(\mathbf{D},\mathbf{E}) = (\mathbf{D}_0,\mathbf{E}_0)\cos\Psi$ inserted, we have

$$\frac{\partial}{\partial t}\iiint(\mathbf{D}\times\mathbf{B})dV = -\oiint d\mathbf{S}\cdot\overleftrightarrow{\mathbf{T}}_M = -\oiint d\mathbf{S}(\mathbf{D}\cdot\mathbf{E})$$
$$= -\iiint\nabla(\mathbf{D}\cdot\mathbf{E})dV = -\iiint(\mathbf{D}_0\cdot\mathbf{E}_0)\sin(2\Psi)(n_d\mathbf{k})dV, \tag{48}$$

resulting in $<\partial/\partial t\iiint(\mathbf{D}\times\mathbf{B})dV>=0$, which means that the EM momentum density is a wave and its distribution varies with space observed at a given time; the EM momentums flowing in and out of a given volume are equal on time average. In other words, there is no momentum transfer taking place between the plane wave and the uniform medium, and thus there is no force acting on the dielectric. This also can be understood through the light-quantized Minkowski's EM-field-cell/photon four-vector $N_p^{-1}(\mathbf{D}\times\mathbf{B},\mathbf{D}\cdot\mathbf{E}/c) = (\hbar n_d\mathbf{k},\hbar\omega/c)$, which characterizes that the momentum of a photon keeps constant during the propagation in the uniform medium. This property can be used to explain why the momentum transfer only takes place on the vacuum-medium interface in the medium Einstein-box thought experiment for a light pulse [40].

The construction of stress tensor is flexible; in a sense, it is artificial. For example, $\nabla\cdot(-\mathbf{DE}-\mathbf{BH})=0$ and $\mathbf{D}\cdot\mathbf{E}=\mathbf{B}\cdot\mathbf{H}$

hold for a plane wave, and the Minkowski's tensor can be re-written in an asymmetric form [23]

$$\overleftrightarrow{\mathbf{T}}_M = -\mathbf{DE} - \mathbf{BH} + \overleftrightarrow{\mathbf{I}}\frac{1}{2}(\mathbf{D}\cdot\mathbf{E} + \mathbf{B}\cdot\mathbf{H}), \tag{49}$$

which does not affect the validity of Eq. (46) and Eq. (48).

Similarly, we can obtain the Abraham's momentum conservation equation, given by

$$\frac{\partial}{\partial t}(\mathbf{E}\times\mathbf{H})/c^2 = -\nabla\cdot\overleftrightarrow{\mathbf{T}}_A, \tag{50}$$

where the Abraham's stress tensor is given by

$$\overleftrightarrow{\mathbf{T}}_A = \beta_{ph}^2[-(\mathbf{ED} + \mathbf{HB}) + \overleftrightarrow{\mathbf{I}}(\mathbf{D}\cdot\mathbf{E})], \tag{51}$$

which is not symmetric. By taking advantage of $\nabla\cdot(\mathbf{DE}) = 0$, $\nabla\cdot(\mathbf{BH}) = 0$, and $\mathbf{D}\cdot\mathbf{E} = \mathbf{B}\cdot\mathbf{H}$, the Abraham's stress tensor can be re-written in a symmetric form, given by

$$\overleftrightarrow{\mathbf{T}}_A = \beta_{ph}^2\left[-(\mathbf{ED} + \mathbf{DE}) - (\mathbf{HB} + \mathbf{BH}) + \overleftrightarrow{\mathbf{I}}\frac{1}{2}(\mathbf{D}\cdot\mathbf{E} + \mathbf{B}\cdot\mathbf{H})\right]. \tag{52}$$

Note that in a moving medium, $\mathbf{ED} = \mathbf{DE}$ and $\mathbf{HB} = \mathbf{BH}$ usually are not true because of the anisotropy of the moving medium. However in free space, $\mathbf{ED} = \mathbf{DE}$ and $\mathbf{HB} = \mathbf{BH}$ always hold because the empty space ($\beta_{ph}^2 = 1$) is isotropic observed in any inertial frames. Thus in free space, $\overleftrightarrow{\mathbf{T}}_M$ given by Eq. (49) and $\overleftrightarrow{\mathbf{T}}_A$ given by Eq. (51) are identical.

It can be seen by comparing Eq. (47) with Eq. (49), and Eq. (51) with Eq. (52) that the symmetry has nothing special in physics: a symmetric form can be turned into asymmetric by adding something, and vice versa.

It also can be seen from Eq. (46) and Eq. (50) that the momentum conservation equations are all differential equations and they can be converted from each other through Maxwell equations. In fact, Eq. (46) and Eq. (50) can be obtained from a more general momentum conservation equation given by $\partial\mathbf{g}/\partial t + \nabla\cdot\overleftrightarrow{\mathbf{T}} = 0$, where $\mathbf{g} = a\mathbf{g}_A + (1-a)\mathbf{g}_M$ and $\overleftrightarrow{\mathbf{T}} = a\overleftrightarrow{\mathbf{T}}_A + (1-a)\overleftrightarrow{\mathbf{T}}_M$, with $\mathbf{g}_A = (\mathbf{E}\times\mathbf{H})/c^2$, $\mathbf{g}_M = \mathbf{D}\times\mathbf{B}$, and $a$ being any constant. We have Eq. (50) for $a = 1$, and Eq. (46) for $a = 0$, while $\mathbf{g}$ is restored to $\mathbf{g} = \mathbf{g}_A = \mathbf{g}_M$ in free space. Just as indicated at the beginning of the paper, Maxwell equations themselves support various forms of momentum conservation equations, resulting in an indeterminacy of momentum definitions. Thus to identify the correctness of momentum definitions, Fermat's principle and the principle of relativity are indispensable tools.

**Unconventional phenomenon for a superluminal medium.** In addition to the negative-frequency appearance, as indicated by Huang [45], a Negative Energy Density (NED) may result for a plane wave when the medium moves opposite to the wave vector direction at a faster-than-dielectric light speed; this phenomenon is called "NED zone" for the sake of convenience.

In the NED zone, the photons possess negative energy from the viewpoint of phenomenological quantum-electrodynamics [54]. The origin of negative EM energy density can be seen from Eq. (43). The energy density Doppler factor is equal to the product of frequency's and EM-field-cell density's. The EM-field-cell density Doppler factor is always positive while the frequency's is negative in the NED zone ($n'_d\hat{\mathbf{n}}'\cdot\boldsymbol{\beta}' = n'_d|\boldsymbol{\beta}'| > 1$), leading to the holding of $\mathbf{D}\cdot\mathbf{E} < 0$ when $\omega < 0$.

In the NED zone, $\mathbf{D}//\mathbf{E}$ and $\mathbf{H}//\mathbf{B}$ are valid, and from Eqs. (20) and (21) we have

$$\frac{\mathbf{D}}{\mathbf{E}} = \frac{1 - |\boldsymbol{\beta}'|/n'_d}{1 - n'_d|\boldsymbol{\beta}'|}\varepsilon' < 0, \tag{53}$$

$$\frac{\mathbf{B}}{\mathbf{H}} = \frac{1 - |\boldsymbol{\beta}'|/n'_d}{1 - n'_d|\boldsymbol{\beta}'|}\mu' < 0. \tag{54}$$

Because of $\omega < 0$ in the NED zone, from Eq. (10-1) we have $\boldsymbol{\beta}_{ph}c = \beta_{ph}c\,\hat{\mathbf{n}}$ with $\beta_{ph}c < 0$. From Eqs. (33) and (34) we have

$$\mathbf{E} = -\beta_{ph}c\,\hat{\mathbf{n}}\times\mathbf{B}, \tag{55}$$

$$\mathbf{H} = +\beta_{ph}c\,\hat{\mathbf{n}}\times\mathbf{D}. \tag{56}$$

It follows that the plane wave in the NED zone is a left-hand wave: (1) $(\mathbf{E}, \mathbf{B}, \hat{\mathbf{n}})$ and $(\mathbf{D}, \mathbf{H}, \hat{\mathbf{n}})$ follow the left-hand rule, and (2) the phase velocity or group velocity is opposite to the wave vector. In other words, the moving medium in the NED zone behaves as a so-called "negative index medium" [41-44], where the refractive index is taken to be negative [42], instead of the frequency here.

As mentioned above, in the NED zone the group velocity $\mathbf{v}_{gr}$ is opposite to the wave vector $n_d\mathbf{k}$, and $\mathbf{D}\cdot\mathbf{E} < 0$ holds. Thus from Eqs. (36)-(38), with $\mathbf{S}_{pseu} = 0$ taken into account, we find that the Poynting vector $\mathbf{S} = \mathbf{S}_{power} = (\mathbf{D}\cdot\mathbf{E})\,\mathbf{v}_{gr}$ also has the same direction as the wave vector $n_d\mathbf{k}$ or the EM momentum $\mathbf{D}\times\mathbf{B}$ in the "negative index medium".

As we have known, the NED zone results from $\omega < 0$. In the NED zone we have $n_d\mathbf{k} = (1 - n'^{-1}_d\boldsymbol{\beta}'\cdot\hat{\mathbf{n}}')\gamma(n'_d\mathbf{k}')$, leading to $(n_d\mathbf{k})//(n'_d\mathbf{k}')$ with $1 - n'^{-1}_d\boldsymbol{\beta}'\cdot\hat{\mathbf{n}}' > 0$. Thus observed in the lab and medium-rest frames respectively, the both power flows have the same direction. From Eqs. (53) and (54) we see that, the moving medium in such a case physically behaves as a "negative index medium". Thus the negative-frequency effect denotes a distinct physical phenomenon where the EM wave is a left-hand wave. In other words, the frequency (EM energy) sign only characterizes the propagation property of EM waves. Experimentally, the observed frequency is always positive, and a positive EM energy propagates along the wave vector direction, while the sign of the frequency is determined by examining the property of wave propagation in the moving medium: (-) for the left-hand and (+) for the right-hand.

It is interesting to point out that, in the effect of "negative index medium" analyzed above, the dispersion of the medium material is not required, and the Poynting vector and EM momentum have the same direction. In contrast, in the traditional effect of negative index medium, first analyzed by Veselago, the medium material must be dispersive to support a positive EM energy, and the Poynting vector is directed opposite to the EM momentum or wave vector [41].

## V. CONCLUSIONS AND REMARKS

By analysis of a plane wave in a moving uniform medium we have shown that, (1) Minkowski's light momentum and energy constitute a Lorentz four-vector, while Abraham's momentum and energy do not, and (2) observed in any inertial frames, Minkowski's EM momentum $\mathbf{g}_M = \mathbf{D}\times\mathbf{B}$ always take the direction of the wave vector $n_d\mathbf{k}$, while the Abraham's momentum $\mathbf{g}_A = \mathbf{E}\times\mathbf{H}/c^2$ does not, unless in free space or when the dielectric medium moves parallel to the wave vector, as shown in Eqs. (35) and (36). The Minkowski's momentum is completely consistent with Fermat's principle and the principle of relativity, and it is the unique correct light momentum.

The photon momentum-energy four-vector $P^{\mu}=\hbar K^{\mu}$ is constructed based on the wave four-vector combined with Einstein light-quantum hypothesis, while the EM momentum-energy four-vector $\bar{P}^{\mu}=N_p^{-1}(\mathbf{D}\times\mathbf{B},\mathbf{D}\cdot\mathbf{E}/c)$ is constructed based on the Lorentz covariance of EM field-strength tensors $F^{\alpha\beta}(\mathbf{E},\mathbf{B})$ and $G^{\alpha\beta}(\mathbf{D},\mathbf{H})$ (to keep Maxwell equations invariant in form in all inertial frames) [47], where $N_p^{-1}\mathbf{D}\times\mathbf{B}$ and $N_p^{-1}\mathbf{D}\cdot\mathbf{E}$ are, respectively, the momentum and energy for a "single EM-field cell". When Einstein's light-quantum hypothesis $N_p^{-1}\mathbf{D}\cdot\mathbf{E}=\hbar\omega$ is imposed on the latter, the latter is restored to the former, namely $N_p^{-1}(\mathbf{D}\times\mathbf{B},\mathbf{D}\cdot\mathbf{E}/c)$ $=(\hbar n_d\mathbf{k},\hbar\omega/c)$, with the "single EM-field cell" becoming "single photon". Thus the light momentum can be described by single photon's momentum or by EM momentum, as stated in the Introduction section.

The principle of relativity, Fermat's principle, and global momentum-energy conservation law are all basic postulates in physics. In the principle-of-relativity frame, it is the Fermat's principle that requires the correct light momentum and energy to constitute a Lorentz four-vector for a plane wave in a moving uniform medium as shown in this paper, while it is the global momentum-energy conservation law that requires the correct light momentum and energy to constitute a Lorentz four-vector in medium Einstein-box thought experiment as shown in [40]. From this we can conclude that the justification of Minkowski's momentum as the correct light momentum is completely required by the basic postulates in physics. This conclusion has a fundamental significance for guiding, although there is no exact plane wave existing in the real world. For example, the fundamental $\mathrm{TE}_{10}$ mode in a rectangular waveguide fully filled with a uniform dielectric will become a uniform plane wave when the transverse dimension of the waveguide goes to infinity; thus the light-momentum result obtained from the $\mathrm{TE}_{10}$ mode should get back to the plane-wave result when the waveguide dimension is much larger the light wavelength. Another typical example is the EM radiation waves generated by finite-size sources, which also can be approximately taken as plane waves locally when the observer is far away from the sources [55].

It should be emphasized that, the significance of the resolution of the Abraham-Minkowski debate presented in the paper is not just to show the justification of the Minkowski's momentum as the unique correct light momentum. In fact, through seeking the resolution we have clarified and developed some basic concepts and principles in electrodynamics and special relativity, which are outlined below:

(i) Poynting vector may not denote the real power flow in an anisotropic medium (confer Eq. (36)-Eq. (38)) [56];

(ii) the Abraham's momentum postulate, which was proposed by Mansuripur and Zakharian [16], is found to be based on an *artificial* assumption that the Poynting vector always denotes a real power flow in *any* situations;

(iii) the total-momentum model [19], which is widely accepted in the community, is proved to be not compatible with the relativity of principle and the global momentum-energy conservation law [40], which are all fundamental postulates in physics;

(iv) the proof of the correctness of both Abraham's and Minkowski's momentums in the relativistic analysis of Einstein-box thought experiment [22] turns out to be obtained by inappropriately using von Laue's theorem of the dynamics of relativity [57];

(v) there may be *apparent* photon velocity and *apparent* photon displacement in a moving medium (see Fig. 2);

(vi) there is an effect of "negative index medium" in the superluminal medium [see Eq. (53) and Eq. (54)];

(vii) the Planck constant is a Lorentz invariant for a plane wave, which is a strict result of relativity, although the Planck constant as a Lorentz invariant is an *implicit postulate* in Dirac relativistic quantum mechanics [58].

One might ask: (1) Does $\mathbf{D}\times\mathbf{B}$ always correctly denote a EM momentum? (2) Is there such a four-vector with the form of $N_p^{-1}(\mathbf{D}\times\mathbf{B},\mathbf{D}\cdot\mathbf{E}/c)$ for any EM fields? The answer is "not necessarily".

To keep Maxwell equations invariant in form, the EM field-strength tensors $F^{\alpha\beta}(\mathbf{E},\mathbf{B})$ and $G^{\alpha\beta}(\mathbf{D},\mathbf{H})$ must follow Lorentz transformations, namely Eqs. (3-1) and (4-1) [47,48]. Based on the field-strength tensors, various forms of stress-energy tensors have been constructed [5,7,11]. However, not every $\mathbf{D}\times\mathbf{B}$ of these EM fields denotes a real EM momentum, and not every EM field has a momentum-energy four-vector in the form of $N_p^{-1}(\mathbf{D}\times\mathbf{B},\mathbf{D}\cdot\mathbf{E}/c)$. To better understand this, some simple examples, including the classical electron mass-energy paradox, are analyzed below, and finally a general EM momentum definition is proposed.

**Example 1.** Suppose that there is a uniform static electric field $\mathbf{E}\neq 0$ while $\mathbf{B}=0$ in free space, with $\mathbf{D}\times\mathbf{B}=0$ and $\mathbf{B}\cdot\mathbf{H}=0$ but $\mathbf{D}\cdot\mathbf{E}\neq 0$, observed in the lab frame *XYZ* (confer Fig. 1). Observed in the moving frame $X'Y'Z'$, from (reversed) EM-field Lorentz transformation Eqs. (3-1) and (4-1) we obtain

$$\mathbf{E}'=\gamma\mathbf{E}-(\gamma-1)\beta^{-2}(\boldsymbol{\beta}\cdot\mathbf{E})\boldsymbol{\beta}, \tag{57}$$

$$\mathbf{B}'=-\gamma\boldsymbol{\beta}\times\mathbf{E}/c=\boldsymbol{\beta}'\times\mathbf{E}'/c\,. \tag{58}$$

Abraham's and Minkowski's momentum are equal, given by

$$\mathbf{D}'\times\mathbf{B}'=(\mathbf{D}'\cdot\mathbf{E}'/c)\boldsymbol{\beta}'-(\mathbf{D}'\cdot\boldsymbol{\beta}')\mathbf{E}'/c\,, \tag{59}$$

which includes two terms, with one in $\boldsymbol{\beta}'-$direction and the other in $\mathbf{E}'-$direction. However the static $\mathbf{E}$-energy fixed in $XYZ$ moves at $\boldsymbol{\beta}'c$ observed in $X'Y'Z'$, and the EM momentum should be parallel to $\boldsymbol{\beta}'$. From this we judge that the term $-(\mathbf{D}'\cdot\boldsymbol{\beta}')\mathbf{E}'/c$ is an *apparent pseudo-momentum*. Especially, when $\boldsymbol{\beta}\,//\,\mathbf{E}$ and $\mathbf{E}\cdot\boldsymbol{\beta}<0$ hold, leading to $\mathbf{E}'=\mathbf{E}$ and $\boldsymbol{\beta}'\cdot\mathbf{E}'>0$, we have $\mathbf{D}'\times\mathbf{B}'=0$ [16], but there should be a non-zero EM momentum observed in $X'Y'Z'$, because the **E**-field has energy and mass, and it is moving. Thus we conclude that $\mathbf{D}'\times\mathbf{B}'$ does not correctly reflect the EM momentum in such a case.

However, when $\boldsymbol{\beta}\perp\mathbf{E}$ or $\boldsymbol{\beta}\cdot\mathbf{E}=0$ holds, we have $-(\mathbf{D}'\cdot\boldsymbol{\beta}')\mathbf{E}'/c=0$. From Eq. (59) and Eqs. (57-58), we obtain $\mathbf{D}'\times\mathbf{B}'=-\gamma^2(\mathbf{D}\cdot\mathbf{E}/c)\boldsymbol{\beta}$ and $\mathbf{E}'\cdot\mathbf{D}'/c=\gamma^2\mathbf{E}\cdot\mathbf{D}/c$, leading to

$$\mathbf{D}'\times\mathbf{B}'=\left(\frac{\mathbf{E}'\cdot\mathbf{D}'}{c^2}\right)\boldsymbol{\beta}'c\,. \tag{60}$$

A direct check indicates that $N_p'^{-1}(\mathbf{D}'\times\mathbf{B}',\mathbf{D}'\cdot\mathbf{E}'/c)$ automatically satisfies the four-vector Lorentz transformation given by Eqs. (1) and (2) with $N_p'/N_p=\gamma$. Note: Such a four-vector is a "frame-dependent" four-vector, instead of a real four-vector. [A real Lorentz four-vector is frame-*in*dependent because it is always a four-vector no matter in what frames it is observed; the Minkowski's four-vector given by Eq. (40) for a plane wave is a real four-vector.]

Nevertheless, it should be pointed out that, $\mathbf{E}'\cdot\mathbf{D}'$ in Eq. (60) is not equal to the EM energy $W'_{em}=0.5(\mathbf{E}'\cdot\mathbf{D}'+\mathbf{B}'\cdot\mathbf{H}')$ in the static field case (instead of a plane wave), and $\mathbf{D}'\times\mathbf{B}'$ cannot denote a real EM momentum either. In other words, if $\mathbf{D}'\times\mathbf{B}'$ in Eq. (60) were taken to be real EM momentum, then according to the conventional definition of momentum = mass times velocity, $\mathbf{E}'\cdot\mathbf{D}'/c^2$ would have to be EM mass but it is not, which directly contradicts Einstein mass-energy equation.

This example clearly shows that when a static electric field moves, the Abraham/Minkowski's momentum does not reflect a real EM momentum, and it cannot be used to constitute a real momentum-energy four-vector.

To correctly understand the relation between Abraham/Minkowski's momentum and EM energy, the first thing we have to do is to define a real EM momentum.

The definition of EM energy $W_{em}=0.5(\mathbf{E}\cdot\mathbf{D}+\mathbf{B}\cdot\mathbf{H})$ is well established and supposed to be valid in any inertial frames; thus the EM momentum for static fields in free space should be defined as

$$\mathbf{g}_{em}=(W_{em}/c^2)\mathbf{v}_{em}\,, \tag{61}$$

where $\mathbf{v}_{em}$ is the EM-energy propagation velocity. Obviously, this definition is compatible with the plane wave in free space, of which the relation between momentum and mass is also well established. Eq. (61) itself does not involve any calculations of $\mathbf{D}\times\mathbf{B}$, and this definition allows $N_{em}^{-1}(\mathbf{g}_{em},W_{em}/c)$ to be a Lorentz four-vector, where $N_{em}$ is the EM-field-cell number density, and $N_{em}^{-1}\mathbf{g}_{em}$ and $N_{em}^{-1}W_{em}$ are, respectively, the momentum and energy for a single EM-field cell (corresponding to a single photon for the plane light-wave case).

The EM energy density can be written as $W'_{em}=0.5(1+\mathbf{B}'\cdot\mathbf{H}'/\mathbf{E}'\cdot\mathbf{D}')\mathbf{E}'\cdot\mathbf{D}'$. Multiplying the both sides of Eq. (60) with $0.5(1+\mathbf{B}'\cdot\mathbf{H}'/\mathbf{E}'\cdot\mathbf{D}')$, we obtain the *real* EM momentum-mass equation, given by

$$\mathbf{g}'_{em}=\frac{1}{2}\left(1+\frac{\mathbf{B}'\cdot\mathbf{H}'}{\mathbf{E}'\cdot\mathbf{D}'}\right)\mathbf{D}'\times\mathbf{B}'=\left(\frac{W'_{em}}{c^2}\right)\boldsymbol{\beta}'c\,. \tag{62}$$

Thus the part $\mathbf{D}'\times\mathbf{B}'-\mathbf{g}'_{em}\neq 0$ behaves as a pseudo-momentum, since $\mathbf{B}'\cdot\mathbf{H}'/\mathbf{E}'\cdot\mathbf{D}'=\beta^2<1$ holds.

It should be emphasized that, not every $\mathbf{D}'\times\mathbf{B}'$ can be converted to a real EM momentum $\mathbf{g}'_{em}$ like Eq. (62); for example, when $\boldsymbol{\beta}\,//\,\mathbf{E}$ and $\mathbf{E}\cdot\boldsymbol{\beta}<0$, we have $\mathbf{D}'\times\mathbf{B}'=0$ from Eq. (59) as mentioned before, but $W'_{em}\neq 0$ and $\mathbf{v}'_{em}=\boldsymbol{\beta}'c=-\boldsymbol{\beta}c\neq 0$, and thus $\mathbf{g}'_{em}=(W'_{em}/c^2)\mathbf{v}'_{em}\neq 0$ cannot be expressed in terms of $\mathbf{D}'\times\mathbf{B}'=0$. In fact, Eq. (59) directly tells us that $\mathbf{g}'_{em}$ and $\mathbf{D}'\times\mathbf{B}'$ are usually not parallel at all.

**Example 2.** Now let us take a look of a symmetric example. Suppose that there is a uniform static magnetic field $\mathbf{B}\neq 0$ while $\mathbf{E}=0$ in free space. Observed in the lab frame *XYZ*, we have $\mathbf{D}\times\mathbf{B}=0$, $\mathbf{D}\cdot\mathbf{E}=0$, but $\mathbf{B}\cdot\mathbf{H}\neq 0$. In the moving frame $X'Y'Z'$, we have $\mathbf{D}'\times\mathbf{B}'=(\mathbf{B}'\cdot\mathbf{H}'/c)\boldsymbol{\beta}'-(\mathbf{B}'\cdot\boldsymbol{\beta}')\mathbf{H}'/c$ by Lorentz transformation. If $\boldsymbol{\beta}\cdot\mathbf{B}=0$ holds, the apparent pseudo-momentum term $-(\mathbf{B}'\cdot\boldsymbol{\beta}')\mathbf{H}'/c$ disappears. Thus we have $\mathbf{D}'\times\mathbf{B}'=-\gamma^2(\mathbf{B}\cdot\mathbf{H}/c)\boldsymbol{\beta}$ and $\mathbf{B}'\cdot\mathbf{H}'/c=\gamma^2\,\mathbf{B}\cdot\mathbf{H}/c$, leading to

$$\mathbf{D}'\times\mathbf{B}'=\left(\frac{\mathbf{B}'\cdot\mathbf{H}'}{c^2}\right)\boldsymbol{\beta}'c\,. \tag{63}$$

The above $\mathbf{D}'\times\mathbf{B}'$ cannot denote a real EM momentum either because $\mathbf{B}'\cdot\mathbf{H}'$ is not equal to the EM energy $W'_{em}=0.5(\mathbf{E}'\cdot\mathbf{D}'+\mathbf{B}'\cdot\mathbf{H}')$ $=0.5(1+\beta^2)\mathbf{B}'\cdot\mathbf{H}'$. Similarly, $N_p'^{-1}(\mathbf{D}'\times\mathbf{B}',\mathbf{B}'\cdot\mathbf{H}'/c)$ is also a frame-dependent four-vector with $N'_p/N_p=\gamma$. From Eq. (63) a real EM momentum-mass equation, like Eq. (62), can be obtained.

**Example 3.** It should be noted that, the pseudo-momentum not only can appear in the moving frame, but also can appear in the static-field-rest frame if the conventional understanding of EM momentum is taken. Suppose that uniform static EM fields $\mathbf{E}\neq 0$ and $\mathbf{B}\neq 0$ with $\mathbf{D}\times\mathbf{B}\neq 0$ in free space are fixed in the lab frame *XYZ*. According to the definition Eq. (61), the real EM momentum is given by $\mathbf{g}_{em}=0$ observed in the lab frame because the static fields are at rest, leading to $\mathbf{v}_{em}=0$. However from the conventional viewpoint, $\mathbf{D}\times\mathbf{B}\neq 0$ is the EM momentum, which is often perplexing [59-61].

In fact, we have reasons to question the conventional understanding of the construction of EM momentum in the above Example 3, where the static fields $\mathbf{E}\neq 0$ and $\mathbf{B}\neq 0$ are completely independent, without any intrinsic relations through Maxwell equations: $\mathbf{E}$ does not produce $\mathbf{B}$, $\mathbf{B}$ does not produce $\mathbf{E}$, and there is no energy coupling between the two fields. Thus physically it is not justifiable to take the mathematically coupling momentum $\mathbf{D}\times\mathbf{B}\neq 0$ to be the EM momentum. In other words, for static **D**-field and **B**-field, which are not intrinsically related, $\mathbf{D}\times\mathbf{B}$ cannot constitute a correct EM momentum.

From above examples, we can obtain two conclusions: (1) for (non-zero) static field $\mathbf{E}$ or $\mathbf{B}$ fixed in the lab frame *XYZ*, observed in a moving frame $X'Y'Z'$, there are pseudo-momentums appearing, and $\mathbf{D}'\times\mathbf{B}'=\mathbf{E}'\times\mathbf{H}'/c^2$ cannot correctly denote a real EM momentum [confer Eq. (59)]; (2) when the frame moves perpendicularly to the field, observed in the moving frame, $\mathbf{D}'\times\mathbf{B}'=\mathbf{E}'\times\mathbf{H}'/c^2$ can be converted into a real EM momentum [see Eq. (62)]; in such a case, a frame-dependent four-vector $N_p'^{-1}(\mathbf{D}'\times\mathbf{B}',\mathbf{D}'\cdot\mathbf{E}'/c)$ or $N_p'^{-1}(\mathbf{D}'\times\mathbf{B}',\mathbf{B}'\cdot\mathbf{H}'/c)$ can be constructed, although $\mathbf{D}'\times\mathbf{B}'$ does not denote a real EM momentum.

Physically realizable static EM fields are "attached-to-source" fields [16], but mathematically Maxwell equations do support static fields without sources, just like a plane wave in free space.

It should be pointed out that, EM-field Lorentz transformation Eqs. (3-1) and (4-1) are valid for any "non-uniform" static fields. In the above examples, the use of "uniform" fields is just to simplify analysis by excluding possible "interfering factors", such as mechanical structures to support the fields.

**Resolution of the classical electron EM mass-energy paradox.** The problem of the classical electron EM energy and momentum has a long story, and it is generally and comprehensibly reviewed in Feynman Lectures [62]; it is also well presented in a recent publication by Griffiths [63].

Feynman presented a specific calculation of EM energy and momentum for a slowly-moving electron in free-space, to show that the mass-energy relation is given by $U_{elec}=(3/4)m_{\mathbf{D}\times\mathbf{B}}c^2$, where the electron's energy is calculated from $U_{elec}=\int(\mathbf{D}\cdot\mathbf{E}/2)dV$ and the electron's mass is calculated from $m_{\mathbf{D}\times\mathbf{B}}=\left|\int(\mathbf{D}\times\mathbf{B})dV\right|/|\mathbf{v}_{elec}|$ (momentum/velocity) [62]. The "infamous factor" (3/4) makes the mass-energy relation not consistent with Einstein mass-energy equation, which has been thought to be "a serious trouble — the failure of the classical electromagnetic theory" [62]. Below we will show that this "serious trouble" is just a paradox.

The inconsistency between Abraham/Minkowski's momentum and EM energy in Feynman's calculations comes from the fact that for a static field it is impossible to find a moving inertial frame, in which Abraham/Minkowski's momentum does not include pseudo-momentum. For example, even when the apparent pseudo-momentum term $-(\mathbf{D}'\cdot\boldsymbol{\beta}')\mathbf{E}'/c=0$ holds in Eq. (59), the first term $(\mathbf{D}'\cdot\mathbf{E}'/c)\boldsymbol{\beta}'$ still includes pseudo-momentum because $\mathbf{D}'\cdot\mathbf{E}'$ is not a real EM energy density, as mentioned before. In other words, the pseudo-momentum will inevitably be involved when calculating the electron's total EM momentum; thus resulting in this paradox.

This paradox can be eliminated by converting Abraham/Minkowski's momentum into a real EM momentum, as shown in Example 1. To do that, first we set the electron fixed at the origin of the lab frame *XYZ*; the electron's static field $\mathbf{E}(\mathbf{x},t)=\mathbf{E}(\mathbf{x})$ produces both electric and magnetic fields $\mathbf{E}'(\mathbf{x}',t')$ and $\mathbf{B}'(\mathbf{x}',t')$ in a moving frame $X'Y'Z'$. To remove the apparent pseudo-momentum term, the $X'Y'Z'$ frame is set to move *perpendicularly* to $\mathbf{E}(\mathbf{x},t)$ at the time-space point $(\mathbf{x},ct)$. Then by Lorentz transformation we obtain the electron's real momentum $\mathbf{g}'_{em}=0.5(1+\beta^2)\mathbf{D}'\times\mathbf{B}'$ at $(\mathbf{x}',ct')$, satisfying Eq. (62). (Not for any $X'Y'Z'-$moving direction there is a conversion from $\mathbf{D}'\times\mathbf{B}'$ to $\mathbf{g}'_{em}$.) This procedure can be done for $\mathbf{E}(\mathbf{x},t)$ at any time-space points. From Eq. (62), we know that the relation between EM mass density $|\mathbf{g}'_{em}|/|\boldsymbol{\beta}'c|$ and energy density $W'_{em}$ for the classical electron automatically satisfies Einstein mass-energy equation at any *differential domains*, with no paradox existing at all.

In conclusion, Feynman's calculations involve pseudo-momentums, resulting in a coefficient difference from Einstein mass-energy equation, but no contradiction to the relativity.

In fact, directly starting from the mass-energy self-consistent momentum definition Eq. (61) we know the electron's EM mass and energy, $m_{elec}=\int(W_{em}/c^2)dV$ and $U_{elec}=\int W_{em}dV$, automatically satisfy Einstein mass-energy relation in any inertial frames. Nevertheless the momentum and energy, $\mathbf{p}_{elec}=(U_{elec}/c^2)\mathbf{v}_{elec}$ and $U_{elec}$, cannot constitute a Lorentz four-vector, which is shown below.

Suppose that the electron is fixed at the origin of the lab frame *XYZ*. Observed in the $X'Y'Z'$ frame moving at $\boldsymbol{\beta}c$ relatively to the lab frame, the electron's velocity is $\mathbf{v}'_{elec}=-\boldsymbol{\beta}c$. In the lab frame, there is no magnetic field (**B** = 0), and from (reversed) Eq. (3-2) and Eq. (4-2) we obtain the EM energy Lorentz transformation, given by

$$W'_{em}=\gamma^2(1+\beta^2)W_{em}-\gamma^2\varepsilon_0(\boldsymbol{\beta}\cdot\mathbf{E})^2\,, \qquad (64)$$

or

$$U'_{elec}=\gamma(1+\beta^2)U_{elec}-\gamma\varepsilon_0\int(\boldsymbol{\beta}\cdot\mathbf{E})^2\,dV\,. \qquad (65)$$

The relation $\mathbf{D}=\varepsilon_0\mathbf{E}$ is used in obtaining Eq. (64), with $\varepsilon_0$ the vacuum permittivity. For the uniform spherical-shell charge distribution used in Feynman's calculations, we have

$$\varepsilon_0\int(\boldsymbol{\beta}\cdot\mathbf{E})^2\,dV=\frac{2}{3}\beta^2U_{elec}\,,\ \text{and}\ \ U_{elec}=\frac{q^2}{8\pi\varepsilon_0 a}\,, \qquad (66)$$

with $q$ the classical electron's charge and $a$ is its radius. From Eqs. (65) and (66) we have [64]

$$U'_{elec}=\left(1+\frac{1}{3}\beta^2\right)\gamma U_{elec}\,. \qquad (67)$$

$U'_{elec}=\gamma U_{elec}$ is a necessary and sufficient condition for $(\mathbf{p}_{elec},U_{elec}/c)$ to be a four-vector (independent of the $X'Y'Z'-$moving direction or $\boldsymbol{\beta}-$direction ). Thus from Eq. (67) we know that the existence of $\beta^2/3$ makes $U'_{elec}\neq\gamma U_{elec}$ so that $(\mathbf{p}_{elec},U_{elec}/c)$ lost four-vector behavior.

In fact, the conclusion $U'_{elec}\neq\gamma U_{elec}$ is valid for any classical electron models because Eq. (65) holds for any static charge distributions. According to Eq. (65), to make $U'_{elec}=\gamma U_{elec}$ hold for any $\boldsymbol{\beta}-$directions with a given $|\boldsymbol{\beta}|-$value , we must have $\beta^2U_{elec}=\varepsilon_0\int(\boldsymbol{\beta}\cdot\mathbf{E})^2dV$ holding for any $\boldsymbol{\beta}-$directions , where the rest electron's EM energy $U_{elec}=0.5\varepsilon_0\int\mathbf{E}^2dV$ is independent of $\boldsymbol{\beta}$. Obviously, it is impossible except for $\boldsymbol{\beta}=0$ or $\mathbf{E}\equiv0$ ; thus $U'_{elec}=\gamma U_{elec}$ cannot hold.

From above we have known that the classical electron's *total* momentum-energy vector $(\mathbf{p}_{elec},U_{elec}/c)$ is not a four-vector although it automatically satisfies Einstein's mass-energy equation. However, as we have indicated previously, the single EM-field-cell momentum-energy vector $N^{-1}_{em}(\mathbf{g}_{em},W_{em}/c)$ is allowed to be a four-vector, where $\mathbf{g}_{em}=(W_{em}/c^2)\mathbf{v}_{em}$, namely Eq. (61). This conclusion is valid for any static fields in free space, and it is proved below.

The proof actually is to determine the $N_{em}-$satisfying condition for given static fields so that $N^{-1}_{em}(\mathbf{g}_{em},W_{em}/c)$ follows four-vector Lorentz transformation.

Suppose that the static fields are fixed in the lab frame $XYZ$. The EM-field energy velocity is given by $\mathbf{v}_{em}=0$ observed in the lab frame, leading to $\mathbf{g}_{em}=0$, while observed in the moving frame $X'Y'Z'$ the energy velocity is $\mathbf{v}'_{em}=-\boldsymbol{\beta}c=\boldsymbol{\beta}'c$. If $N^{-1}_{em}(\mathbf{g}_{em},W_{em}/c)$ is a four-vector, then it satisfies the Lorentz transformation

$$\frac{\mathbf{g}'_{em}}{N'_{em}}=\frac{\mathbf{g}_{em}}{N_{em}}+\frac{\gamma-1}{\beta^2}\boldsymbol{\beta}\cdot\left(\frac{\mathbf{g}_{em}}{N_{em}}\right)\boldsymbol{\beta}-\gamma\boldsymbol{\beta}\left(\frac{W_{em}}{cN_{em}}\right), \qquad (68)$$

$$\frac{W'_{em}}{cN'_{em}}=\gamma\left[\frac{W_{em}}{cN_{em}}-\boldsymbol{\beta}\cdot\left(\frac{\mathbf{g}_{em}}{N_{em}}\right)\right]. \qquad (69)$$

From above we obtain the necessary and sufficient condition for $N^{-1}_{em}(\mathbf{g}_{em},W_{em}/c)$ to be a four-vector, given by

$$(N'^{-1}_{em}W'_{em})=\gamma(N^{-1}_{em}W_{em})\,,\quad\text{or}\quad \frac{N_{em}}{N'_{em}}=\gamma\frac{W_{em}}{W'_{em}}\,, \qquad (70)$$

which makes Eqs. (68) and (69) hold at the same time. Note that all field quantities must satisfy EM-field strength tensor Lorentz transformations given by Eqs. (3) and (4).

From Eq. (70), we find that the single EM-field-cell energy relation $(N'^{-1}_{em}W'_{em})=\gamma(N^{-1}_{em}W_{em})$ is different from the total energy relation $U'_{elec}=\gamma(1+\beta^2/3)U_{elec}$ for the classical electron given by Eq. (67). It is this difference that results in the difference of Lorentz property between $N^{-1}_{em}(\mathbf{g}_{em},W_{em}/c)$ and $(\mathbf{p}_{elec},U_{elec}/c)$.

Recall the definition of refractive index for a plane wave in a medium, given by $|\text{wave vector}|/|\omega/c|=|n_d\mathbf{k}|/|\omega/c|$. Obviously, the index definition is not applicable to static fields where $n_d\mathbf{k}=0$ and $\omega/c=0$ hold, and the index effect on the momentum-associated mass does not exist; in other words, the momentum-associated mass is equal to the energy-associated mass for any static fields. Therefore, the EM momentum definition $\mathbf{g}_{em}=(W_{em}/c^2)\mathbf{v}_{em}$ given by Eq. (61) and the four-vector behavior of $N^{-1}_{em}(\mathbf{g}_{em},W_{em}/c)$ are valid for *any static fields both in free space and in a medium*.

**General EM momentum definition.** Based on the analysis of various simple but basic forms of EM fields in this paper, it makes sense to generalize the result of Eq. (35), given by $\mathbf{D}\times\mathbf{B}=(n_d/c)^2W_{em}\mathbf{v}_{gr}$ for a plane wave in a medium, as a general EM momentum definition, given by

$$\mathbf{g}_{em}=n^2_{eff}\left(\frac{W_{em}}{c^2}\right)\mathbf{v}_{em}\,, \qquad (71)$$

which has exactly the same form as that of the momentum definition in Newton classical mechanics.

For EM waves, $n_{eff} = |\text{wave vector}|/|\omega/c|$ is an effective refractive index, which has an index effect on the EM momentum-associated mass $\rho_{\mathbf{P}} = n_{eff}^2 (W_{em}/c^2)$. In a regular dielectric-filled metallic waveguide, for example, the effective index is given by $n_{eff} = (n_d^2 - k_c^2 c^2/\omega^2)^{1/2}$ with $k_c$ the cutoff wave number. For a plane wave in a medium (free space), $n_{eff} = n_d$ $(=1)$ holds because of $k_c = 0$. However for general dispersion structures, $0 < n_{eff} < n_d$ holds. That means that the relation between EM momentum and refractive index $n_d$ also depends on the dispersion property of EM structure (geometry). It is possible for different dependences of EM momentum on refractive index $n_d$ to be observed experimentally in different wave-dispersion structures with the *same dielectric material*, or even the same structure but operating at *different frequencies*.

For static fields, $n_{eff} = 1$ is taken no matter whether in vacuum or in a medium, because there is no index effect on EM mass, and the momentum-associated mass equal to the energy-associated mass, namely $\rho_{\mathbf{P}} = \rho_E = W_{em}/c^2$ both in vacuum and in a medium.

There are some points that need to be emphasized. (1) For a plane wave in a medium (free space), $n_{eff} = n_d$ $(=1)$ holds, and Eq. (71) is equivalent to $\mathbf{g}_{em} = \mathbf{D} \times \mathbf{B}$ (Minkowski momentum) because of the EM energy speed $|\mathbf{v}_{em}| = c/n_d$. (2) For static fields fixed in the lab frame $XYZ$, the EM momentum is given by $\mathbf{g}_{em} = 0$ observed in the lab frame because of the energy velocity $\mathbf{v}_{em} = 0$, while observed in the frame $X'Y'Z'$ moving with respect to the lab frame at $\boldsymbol{\beta} c$, the EM momentum is given by $\mathbf{g}'_{em} = (W'_{em}/c^2)\boldsymbol{\beta}' c$, because of the energy velocity $\mathbf{v}'_{em} = \boldsymbol{\beta}' c = -\boldsymbol{\beta} c$. (3) For general EM waves, the momentum definition Eq. (71) is a generalization from the momentum formulation of the plane wave in a medium based on the physical requirement: EM-wave momentum is equal to the momentum-associated mass multiplied by the energy velocity.

Interestingly, the dispersion property of wave momentum may make the momentum itself behaving as "Minkowski-type momentum" or "Abraham-type momentum", depending on the EM structure's operating frequency, if the traditional criterions are employed. Such an example is given below.

Suppose that a travelling wave propagating along the *z*-direction in an above-mentioned regular dielectric-filled metallic waveguide. The energy velocity, equal to the group velocity, is given by $|\mathbf{v}_{em}| = n_{eff} c/n_d^2$, and from Eq. (71) we obtain the wave momentum (in the *z*-direction), given by

$$|\mathbf{g}_{em}| = n_d \left(1 - \frac{\omega_c^2}{\omega^2}\right)^{\frac{3}{2}} \left(\frac{W_{em}}{c}\right), \qquad \text{with } \omega_c = \frac{k_c c}{n_d}, \tag{72}$$

where $k_c$ is a structure's constant required by boundary conditions.

When the operating frequency $\omega$ is much larger than the cutoff frequency $\omega_c$, we have the "Minkowski-type EM momentum"

$$|\mathbf{g}_{em}| \approx n_d \left(\frac{W_{em}}{c}\right), \qquad \text{for } \omega >> \omega_c, \tag{73}$$

which corresponds to the Minkowski's photon momentum formulation $p_M = n_d \hbar\omega/c$ discussed in Sec. III.

When the frequency is set to be $\omega = (n_d \omega_c)(n_d^2 - n_d^{2/3})^{-1/2}$ $(> \omega_c)$, we have the "Abraham-type EM momentum"

$$|\mathbf{g}_{em}| = \frac{1}{n_d}\left(\frac{W_{em}}{c}\right), \qquad \text{for } \omega = \frac{n_d \omega_c}{\sqrt{n_d^2 - n_d^{2/3}}}, \tag{74}$$

which corresponds to the Abraham's photon momentum formulation $p_A = \hbar\omega/(n_d c)$ discussed in Sec. III.

Of course, in general cases neither Abraham-type nor Minkowski-type momentum can be identified in terms of the traditional criterions.

We have shown that the EM momentum and energy for a plane wave, which has no dispersion, constitute a Lorentz four-vector in a form of single photon or EM-field cell. However for dispersive EM waves (interfering waves), that are guided by EM structures with finite dimensions and of which the momentums are tailored to propagate in a required direction, whether and how the momentum and energy can constitute a four-vector is still an open question.

Finally, we would like to make a comment on the Planck constant $\hbar$, which is widely thought to be a "universal constant" [65]. For a plane wave in a moving medium, $(n'_d \mathbf{k}', \omega'/c)$ is Lorentz covariant; thus resulting in the invariance of Planck constant, as shown in Sec. IV in the paper. Mathematically speaking, if $(n'_d \mathbf{k}', \omega'/c)$ is Lorentz covariant, then the invariance of $\hbar$ and the covariance of $\hbar(n'_d \mathbf{k}', \omega'/c)$ are equivalent. However for a moving point light source in free space where $n'_d = 1$ holds, the covariance of $(\mathbf{k}', \omega'/c)$ is destroyed, and the invariance of Planck constant is questioned [55,66].

## Attachment-I. Comparison of results from Gordon-metric and Minkowski-metric dispersion relations

As we know, a correct physical result does not depend on the mathematical method that is used. Eq. (8) in this paper, resulting in an invariant form of refractive index, is a Minkowski-metric dispersion relation, because it is directly obtained from the Minkowski-metric-expressed Lorentz scalar equation $g^{\mu\nu}(K_\mu K_\nu - K'_\mu K'_\nu) = 0$ based on the wave four-vector $K^\mu = (n_d \mathbf{k}, \omega/c)$. However, some scientists insist that,

> "The dispersion relation (8), which underlies many computations in this paper, is incorrect. Instead of the Minkowski spacetime metric, one should use the so-called optical (or Gordon) metric, and the correct dispersion relation then reads as the Eq. (A7) in T. Ramos, G. F. Rubilar, Y. N. Obukhov, Phys. Lett. A **375** (2011) 1703." (**RRO-paper** for short: http://arxiv.org/abs/1103.1654; 10.1016/j.physleta.2011.03.015 )

Since a correct physical result does not depend on mathematical methods, we can affirm that the results respectively from Minkowski- and Gordon-metric dispersion relations must be the same. In this attachment, we will show that the result from Eq. (8) in this paper is, indeed, the same as that from the Gordon-metric-expressed dispersion relation Eq. (A7) given in **RRO-paper**. [In the next attachment, we will show that Eq. (8) is exactly equivalent to Eq. (A7) of **RRO-paper**.]

To convince readers, a specific proof is given below. For readers' convenience, the same math symbols as those used in **RRO-paper** are adopted below.

Eq. (A7) of **RRO-paper** (http://arxiv.org/abs/1103.1654 ) reads

$$\gamma^{\mu\nu} k_\mu k_\nu = 0, \tag{PRO-A7}$$

where $k_\mu$ is the wave 4-vector in the *Lab frame*, given by

$$k_\mu = (\omega, -k, 0, 0), \tag{PRO-A8}$$

namely

$$k_1 = \omega, \qquad k_2 = -k, \qquad k_3 = 0, \qquad k_4 = 0,$$

and $\gamma^{\mu\nu}$ is the so-called Gordon optical tensor, given by

$$\gamma^{\mu\nu} = \begin{pmatrix} \frac{1}{c^2}[1+(n^2-1)\gamma^2] & \frac{1}{c}[(n^2-1)\gamma^2\beta] & 0 & 0 \\ \frac{1}{c}[(n^2-1)\gamma^2\beta] & -1+(n^2-1)\gamma^2\beta^2 & 0 & 0 \\ 0 & 0 & -1 & 0 \\ 0 & 0 & 0 & -1 \end{pmatrix}, \tag{PRO-A9}$$

namely

$$\gamma^{11} = \frac{1}{c^2}[1+(n^2-1)\gamma^2], \qquad \gamma^{12} = \frac{1}{c}[(n^2-1)\gamma^2\beta], \qquad \gamma^{13} = 0, \qquad \gamma^{14} = 0,$$

$$\gamma^{21} = \gamma^{12} = \frac{1}{c}[(n^2-1)\gamma^2\beta], \qquad \gamma^{22} = -1+(n^2-1)\gamma^2\beta^2, \qquad \gamma^{23} = 0, \qquad \gamma^{24} = 0$$

$$\gamma^{31} = 0, \qquad \gamma^{32} = 0, \qquad \gamma^{33} = -1, \qquad \gamma^{34} = 0,$$

$$\gamma^{41} = 0, \qquad \gamma^{42} = 0, \qquad \gamma^{43} = 0, \qquad \gamma^{44} = -1.$$

In the above, $\beta c$ is the $x$-direction moving velocity of the medium (only 1D motion is assumed in **PRO-paper**), and $\gamma = (1-\beta^2)^{-1/2}$. $\omega$ and $k$, are, respectively the frequency and the wave number in the *lab* frame, but $n$ is the refractive index in the *medium-rest* frame.

From Eq. (PRO-A7), we have

$$\gamma^{\mu\nu} k_\mu k_\nu = \gamma^{11} k_1 k_1 + \gamma^{12} k_1 k_2 + \gamma^{21} k_2 k_1 + \gamma^{22} k_2 k_2 + \gamma^{33} k_3 k_3 + \gamma^{44} k_4 k_4 = 0$$

namely

$$\gamma^{\mu\nu} k_\mu k_\nu = \frac{1}{c^2}[1+(n^2-1)\gamma^2]\omega^2 + 2\frac{1}{c}[(n^2-1)\gamma^2\beta](-k)\omega + [-1+(n^2-1)\gamma^2\beta^2](-k)^2 = 0. \tag{I-1}$$

From above Eq. (I-1), we have

$$kc = \frac{(n^2-1)\gamma^2\beta - n}{[-1+(n^2-1)\gamma^2\beta^2]}\omega \,. \tag{I-2}$$

Since

$$\text{numerator} = (n^2-1)\gamma^2\beta - n = (n^2-1)\gamma^2\beta - n\gamma^2(1-\beta^2) = -\gamma^2(1-n\beta)(n+\beta)\,,$$

$$\text{denominator} = -1+(n^2-1)\gamma^2\beta^2 = -\gamma^2(1-\beta^2)+(n^2-1)\gamma^2\beta^2 = -\gamma^2(1+n\beta)(1-n\beta)\,,$$

we have

$$kc = \frac{n+\beta}{1+n\beta}\omega\,, \qquad \text{namely} \qquad \gamma^{\mu\nu}k_\mu k_\nu = 0 \;\Rightarrow\; |kc| = \left|\frac{n+\beta}{1+n\beta}\omega\right|. \tag{I-3}$$

Now let us convert the symbols in the above Eq. (I-3) into those used in this paper:

$$n_d|\mathbf{k}| = n_d\frac{|\omega|}{c} \to |k| \qquad\qquad n'_d \to n\,, \tag{I-4}$$

we have

$$n_d\frac{|\omega|}{c}c = \left|\frac{n'_d+\beta}{1+n'_d\beta}\omega\right|,$$

or

$$n_d = \left|\frac{n'_d+\beta}{1+n'_d\beta}\right|. \tag{I-5}$$

The above Eq.(I-5) is right the Eq. (15) in this paper.

In conclusion, the result from Eq. (A7) derived from so-called Gordon optical metric in **RRO-paper** (http://arxiv.org/abs/1103.1654) is the same as that from Eq. (8) in this paper.

## Attachment-II. Equivalence of Minkowski-metric and Gordon-metric dispersion relations

In this attachment, we will show that the Minkowski-metric and Gordon-metric dispersion relations are equivalent. As shown in Sec. II, Eq. (8) is directly obtained from the Minkowski-metric dispersion equation, given by

$$g^{\mu\nu}(K_\mu K_\nu - K'_\mu K'_\nu) = 0\,, \tag{II-1}$$

where $K'_\mu = (-n'_d\mathbf{k}', \omega'/c) = g_{\mu\nu}K'^\nu$, with $n'_d = \sqrt{\mu'\varepsilon'/\mu_0\varepsilon_0}$ the refractive index in the medium-rest frame, $K'^\mu = (n'_d\mathbf{k}', \omega'/c)$, Minkowski metric $g_{\mu\nu} = g^{\mu\nu} = diag(-1,-1,-1,+1)$, and $|\mathbf{k}'| = |\omega'/c|$ resulting from wave equation for a uniform plane wave. Note: $K'^4 = K'_4 = \omega'/c$. In the Minkowski space, the contra-variant coordinate is given by $X^\mu = (x, y, z, ct)$, with

$$dX'^\mu = \frac{\partial X'^\mu}{\partial X^\lambda}dX^\lambda \qquad \text{(contra-variant)}, \tag{II-2}$$

and

$$\frac{\partial}{\partial X'^\mu} = \frac{\partial X^\lambda}{\partial X'^\mu}\frac{\partial}{\partial X^\lambda} \qquad \text{(covariant)}. \tag{II-3}$$

With $K'_4 = \omega'/c$ taken into account, we have

$$g^{\mu\nu}K'_\mu K'_\nu = \left(\frac{\omega'}{c}\right)^2(1-n'^2_d) = (1-n'^2_d)(K'_4)^2 \,. \tag{II-4}$$

Sine $K'_4$ is a covariant component, transforming like Eq. (II-3), we have

$$(K'_4)^2 = \left(\frac{\partial X^\mu}{\partial X'^4}K_\mu\right)^2 = \frac{\partial X^\mu}{\partial X'^4}\frac{\partial X^\nu}{\partial X'^4}K_\mu K_\nu \,. \tag{II-5}$$

Inserting Eq. (II-5) into Eq. (II-4), we have

$$g^{\mu\nu}K'_\mu K'_\nu = (1-n'^2_d)\frac{\partial X^\mu}{\partial X'^4}\frac{\partial X^\nu}{\partial X'^4}K_\mu K_\nu \,. \tag{II-6}$$

Inserting Eq. (II-6) into Eq. (II-1), we have

$$\left(g^{\mu\nu} - (1-n'^2_d)\frac{\partial X^\mu}{\partial X'^4}\frac{\partial X^\nu}{\partial X'^4}\right)K_\mu K_\nu = 0 \,. \tag{II-7}$$

From above, we obtain the Gordon optical metric, given by

$$\Gamma^{\mu\nu} = g^{\mu\nu} - (1-n'^2_d)\frac{\partial X^\mu}{\partial X'^4}\frac{\partial X^\nu}{\partial X'^4} \,. \tag{II-8}$$

Thus the Gordon-metric dispersion relation Eq. (II-7) can be re-written as

$$\Gamma^{\mu\nu}K_\mu K_\nu = 0 \,. \tag{II-9}$$

Now let us determine all the components of $\Gamma^{\mu\nu}$. Since $X^\mu = (\mathbf{x}, ct)$, from Eqs. (1) and (2) in Sec. II we have

$$X^1 = \left(\mathbf{x}' + \frac{\gamma-1}{\beta^2}(\boldsymbol{\beta}'\cdot\mathbf{x}')\boldsymbol{\beta}' - \gamma\boldsymbol{\beta}'ct'\right)\cdot\hat{\mathbf{x}} \,, \tag{II-10}$$

$$X^2 = \left(\mathbf{x}' + \frac{\gamma-1}{\beta^2}(\boldsymbol{\beta}'\cdot\mathbf{x}')\boldsymbol{\beta}' - \gamma\boldsymbol{\beta}'ct'\right)\cdot\hat{\mathbf{y}} \,, \tag{II-11}$$

$$X^3 = \left(\mathbf{x}' + \frac{\gamma-1}{\beta^2}(\boldsymbol{\beta}'\cdot\mathbf{x}')\boldsymbol{\beta}' - \gamma\boldsymbol{\beta}'ct'\right)\cdot\hat{\mathbf{z}} \,, \tag{II-12}$$

$$X^4 = ct = \gamma(ct' - \boldsymbol{\beta}'\cdot\mathbf{x}') \,, \tag{II-13}$$

where $\hat{\mathbf{x}}$, $\hat{\mathbf{y}}$, and $\hat{\mathbf{z}}$ are, respectively, the unit vectors of three frame's axes.

From Eqs. (II-10)-(II-13), we obtain

$$\frac{\partial X^1}{\partial X'^4} = \frac{\partial X^1}{\partial(ct')} = -\gamma\beta'_x = \gamma\beta_x \,, \tag{II-14}$$

$$\frac{\partial X^2}{\partial X'^4} = \frac{\partial X^2}{\partial(ct')} = -\gamma\beta'_y = \gamma\beta_y \,, \tag{II-15}$$

$$\frac{\partial X^3}{\partial X'^4} = \frac{\partial X^3}{\partial(ct')} = -\gamma\beta'_z = \gamma\beta_z \,, \tag{II-16}$$

$$\frac{\partial X^4}{\partial X'^4}=\frac{\partial X^4}{\partial (ct')}=\gamma \,. \tag{II-17}$$

The above Eqs. (II-14)-(II-17) can be simply written as $\partial X^\mu / \partial X'^4 = (\gamma\boldsymbol{\beta},\gamma)$ , which is a four-vector.

Inserting Eqs. (II-14)-(II-17) into Eq. (II-8), we obtain the Gordon-metric matrix

$$\Gamma^{\mu\nu}=\begin{pmatrix}-1 & 0 & 0 & 0\\ 0 & -1 & 0 & 0\\ 0 & 0 & -1 & 0\\ 0 & 0 & 0 & 1\end{pmatrix}-\gamma^2(1-n_d'^2)\begin{pmatrix}\beta_x^2 & \beta_x\beta_y & \beta_x\beta_z & \beta_x\\ \beta_y\beta_x & \beta_y^2 & \beta_y\beta_z & \beta_y\\ \beta_z\beta_x & \beta_z\beta_y & \beta_z^2 & \beta_z\\ \beta_x & \beta_y & \beta_z & 1\end{pmatrix}. \tag{II-18}$$

By setting $B^\mu = \partial X^\mu / \partial X'^4 = (\gamma\boldsymbol{\beta},\gamma)$ , both Eq. (II-8) and Eq. (II-18) can be re-written as

$$\Gamma^{\mu\nu}=g^{\mu\nu}-(1-n_d'^2)B^\mu B^\nu \,. \tag{II-19}$$

Obviously, the above derivations are reversible.

For the medium moving only along the *x*-direction ( $\beta_x=\beta,\ \beta_y=\beta_z=0$ ), we have a simplified Gordon metric, given by

$$\Gamma^{\mu\nu}=\begin{pmatrix}-1-(1-n_d'^2)\gamma^2\beta^2 & 0 & 0 & -(1-n_d'^2)\gamma^2\beta\\ 0 & -1 & 0 & 0\\ 0 & 0 & -1 & 0\\ -(1-n_d'^2)\gamma^2\beta & 0 & 0 & 1-(1-n_d'^2)\gamma^2\end{pmatrix}. \tag{II-20}$$

Note: Because the definitions of Minkowski metric and the wave four-vector are different from those used in **RRO-paper** (http://arxiv.org/abs/1103.1654), there are some differences in component orders and *c*-factor in Eq. (II-20), compared with Eq. (A9) of **RRO-paper**.

**Conclusion:** The Minkowski-metric dispersion relation Eq. (II-1) or Eq. (8) in Sec. II is equivalent to the Gordon-optical-metric dispersion relation Eq. (II-9), or Eq. (A7) given in **RRO-paper** (http://arxiv.org/abs/1103.1654). However, it is interesting to point out that, this Minkowski-metric dispersion relation has never been realized in the community although its derivation is so simple and straightforward.

**Derivation summary.** The derivation for the equivalence of Gordon-metric and Minkowski-metric dispersion equations can be summarized as follow:

$$g^{\mu\nu}(K_\mu K_\nu - K'_\mu K'_\nu)=0 \qquad \text{(Minkowski-metric dispersion relation)} \tag{II-21}$$

$$\begin{aligned}\Leftrightarrow\ g^{\mu\nu}K_\mu K_\nu = g^{\mu\nu}K'_\mu K'_\nu &= (1-n_d'^2)\left(\frac{\omega'}{c}\right)^2 = (1-n_d'^2)(K'_4)^2\\ &= (1-n_d'^2)\left(\frac{\partial X^\mu}{\partial X'^4}K_\mu\right)^2\\ &= (1-n_d'^2)(B^\mu K_\mu)^2\\ &= (1-n_d'^2)B^\mu B^\nu K_\mu K_\nu\end{aligned} \tag{II-22}$$

$$\Leftrightarrow\ \left[g^{\mu\nu}-(1-n_d'^2)B^\mu B^\nu\right]K_\mu K_\nu = 0 \tag{II-23}$$

$$\Leftrightarrow\ \Gamma^{\mu\nu}K_\mu K_\nu = 0 \qquad \text{(Gordon-metric dispersion relation)}$$

Namely

$$g^{\mu\nu}(K_\mu K_\nu - K'_\mu K'_\nu) = 0 \qquad \Leftrightarrow \qquad \Gamma^{\mu\nu} K_\mu K_\nu = 0\,. \tag{II-24}$$

**Comparison of dispersion equations.** What difference between $g^{\mu\nu}(K_\mu K_\nu - K'_\mu K'_\nu) = 0$ and $\Gamma^{\mu\nu} K_\mu K_\nu = 0$ ?

(1) Minkowski's: $g^{\mu\nu}(K_\mu K_\nu - K'_\mu K'_\nu) = 0$

$$\Rightarrow \quad \left(\frac{\omega}{c}\right)^2 (1 - n_d^2) = \left(\frac{\omega'}{c}\right)^2 (1 - n_d'^2)$$

(i) if using Doppler-condition $\dfrac{\omega}{c} = \gamma\left(\dfrac{\omega'}{c} - n'_d\, \mathbf{k}' \cdot \boldsymbol{\beta}'\right)$,

$$\Rightarrow \quad \left[\gamma\left(\frac{\omega'}{c} - n'_d\, \mathbf{k}' \cdot \boldsymbol{\beta}'\right)\right]^2 (1 - n_d^2) = \left(\frac{\omega'}{c}\right)^2 (1 - n_d'^2)$$

$$\Rightarrow \quad n_d = \frac{\sqrt{(n_d'^2 - 1) + \gamma^2 (1 - n'_d \hat{\mathbf{n}}' \cdot \boldsymbol{\beta}')^2}}{\left|\gamma (1 - n'_d \hat{\mathbf{n}}' \cdot \boldsymbol{\beta}')\right|} \qquad [\,\omega' > 0 \text{ assumed; see Eq. (9)}]. \tag{II-25-1}$$

(ii) if using Doppler-condition $\dfrac{\omega'}{c} = \gamma\left(\dfrac{\omega}{c} - n_d\, \mathbf{k} \cdot \boldsymbol{\beta}\right)$,

$$\Rightarrow \quad \left(\frac{\omega}{c}\right)^2 (1 - n_d^2) = \left[\gamma\left(\frac{\omega}{c} - n_d\, \mathbf{k} \cdot \boldsymbol{\beta}\right)\right]^2 (1 - n_d'^2)$$

$$\Rightarrow \quad n'_d = \frac{\sqrt{(n_d^2 - 1) + \gamma^2 (1 \mp n_d \hat{\mathbf{n}} \cdot \boldsymbol{\beta})^2}}{\left|\gamma (1 \mp n_d \hat{\mathbf{n}} \cdot \boldsymbol{\beta})\right|} \qquad \text{with } \begin{cases} - \text{ sign} \rightarrow \omega > 0 \\ + \text{ sign} \rightarrow \omega < 0 \end{cases}. \tag{II-25-2}$$

(2) Gordon's: $\Gamma^{\mu\nu} K_\mu K_\nu = 0 \quad \Leftrightarrow \quad g^{\mu\nu} K_\mu K_\nu = \left(\dfrac{\omega'}{c}\right)^2 (1 - n_d'^2) = \left(\dfrac{\partial X^\mu}{\partial X'^4} K_\mu\right)^2 (1 - n_d'^2)$ from Eq. (II-22)

$$\Rightarrow \quad \left(\frac{\omega}{c}\right)^2 (1 - n_d^2) = \left(\frac{\partial X^\mu}{\partial X'^4} K_\mu\right)^2 (1 - n_d'^2)$$

$$\Rightarrow \quad \left(\frac{\omega}{c}\right)^2 (1 - n_d^2) = \left[\gamma\left(\frac{\omega}{c} - n_d\, \mathbf{k} \cdot \boldsymbol{\beta}\right)\right]^2 (1 - n_d'^2) \quad \text{since } \frac{\partial X^\mu}{\partial X'^4} K_\mu = \frac{\omega'}{c} = \gamma\left(\frac{\omega}{c} - n_d\, \mathbf{k} \cdot \boldsymbol{\beta}\right), \text{ Doppler-condition}$$

$$\Rightarrow \quad n'_d = \frac{\sqrt{(n_d^2 - 1) + \gamma^2 (1 \mp n_d \hat{\mathbf{n}} \cdot \boldsymbol{\beta})^2}}{\left|\gamma (1 \mp n_d \hat{\mathbf{n}} \cdot \boldsymbol{\beta})\right|} \qquad \text{with } \begin{cases} - \text{ sign} \rightarrow \omega > 0 \\ + \text{ sign} \rightarrow \omega < 0 \end{cases}. \tag{II-26}$$

From above, we see that $g^{\mu\nu}(K_\mu K_\nu - K'_\mu K'_\nu) = 0$ itself does not include Doppler-condition, while $\Gamma^{\mu\nu} K_\mu K_\nu = 0$ has already included a specific Doppler-condition. Eq. (II-26) and Eq. (II-25-2) are the same because the same Doppler-condition is included.

$\Gamma^{\mu\nu} K_\mu K_\nu = 0$ looks simple, but one has to do much more calculations to get the specific dispersion relation from $\Gamma^{\mu\nu}$ given in Eq. (II-18).

## Attachment-III. Is a tensor form of physical law guaranteed to be its compatibility with special principle of relativity?

It is well recognized in the community that all physical laws expressed in a tensor form are guaranteed to be compatible with the principle of relativity. However, this "common sense" is challenged by the light particle (photon) 4-velocity vector (first-rank tensor) in a moving medium, which is widely presented in electrodynamics textbooks and literature to illustrate relativistic velocity addition rule through Fizeau experiment. [For example, see: W. Pauli, Theory of relativity, (Pergamon Press, London, 1958), Eq. (14), p. 18, Sec. 6.] As shown below, this light particle 4-velocity is not compatible with the relativity principle, because the obtained light particle velocity observed in the lab frame does not has the same direction as the light wave vector has, unless the medium moves parallel to the wave vector (the Fizeau-experiment case).

Suppose that the light particle 4-velocity in the medium-rest frame is given by $U'^{\mu} = \gamma'_{lt}(\mathbf{u}'_{lt}, c)$, where $|\mathbf{u}'_{lt}| = c/n'_d$ is the dielectric light speed, with $n'_d$ the refractive index, $c$ the light speed in free space, $\gamma'_{lt} = (1-\mathbf{u}'^2_{lt}/c^2)^{-1/2}$, and $g_{\mu\nu}U'^{\mu}U'^{\nu} = \gamma'^2_{lt}(c^2 - \mathbf{u}'^2_{lt}) = c^2$. Observed in the medium-rest frame, $\mathbf{u}'_{lt}$ has the same direction as the wave vector $n'_d\mathbf{k}' = (n'_d|\mathbf{k}'|)\hat{\mathbf{n}}'$ has, namely $\mathbf{u}'_{lt} = |\mathbf{u}'_{lt}|\hat{\mathbf{n}}' = c/n'_d\,\hat{\mathbf{n}}'$. From the Lorentz transformation given by Eqs. (1) and (2) in Sec. II, we obtain the light particle velocity $\mathbf{u}_{lt}$ in the lab frame, given through

$$\gamma_{lt}\mathbf{u}_{lt} = (\gamma'_{lt}\mathbf{u}'_{lt}) + \frac{\gamma-1}{\beta^2}\boldsymbol{\beta}'\cdot(\gamma'_{lt}\mathbf{u}'_{lt})\boldsymbol{\beta}' - \gamma\boldsymbol{\beta}'(\gamma'_{lt}c)\,, \qquad \text{with } \gamma_{lt} = (1-\mathbf{u}^2_{lt}/c^2)^{-1/2}. \tag{III-1}$$

$$\gamma_{lt}c = \gamma[\gamma'_{lt}c - \boldsymbol{\beta}'\cdot(\gamma'_{lt}\mathbf{u}'_{lt})]\,. \tag{III-2}$$

From above Eq. (III-1) and Eq. (7) in Sec. II, we have

$$\gamma_{lt}\mathbf{u}_{lt} = (\gamma'_{lt}|\mathbf{u}'_{lt}|)\hat{\mathbf{n}}' + \left[\frac{\gamma-1}{\beta^2}(\boldsymbol{\beta}'\cdot\hat{\mathbf{n}}')(\gamma'_{lt}|\mathbf{u}'_{lt}|) - \gamma(\gamma'_{lt}c)\right]\boldsymbol{\beta}'\,, \tag{III-3}$$

$$n_d\mathbf{k} = (n'_d|\mathbf{k}'|)\hat{\mathbf{n}}' + \left[\frac{\gamma-1}{\beta^2}(\boldsymbol{\beta}'\cdot\hat{\mathbf{n}}')(n'_d|\mathbf{k}'|) - \gamma\left(\frac{\omega'}{c}\right)\right]\boldsymbol{\beta}'\,. \tag{III-4}$$

If $\boldsymbol{\beta}' // \hat{\mathbf{n}}'$ (including $\boldsymbol{\beta}' = 0$) is valid, observed in the lab frame, the light particle velocity is apparently parallel to the wave vector or $\mathbf{u}_{lt} // (n_d\mathbf{k})$. If $\hat{\mathbf{n}}'$ and $\boldsymbol{\beta}'$ are linearly independent ($\boldsymbol{\beta}' // \hat{\mathbf{n}}'$ not hold), a sufficient and necessary condition for $\mathbf{u}_{lt} // (n_d\mathbf{k})$ is that their coefficient determinant is equal to zero, namely

$$\begin{vmatrix} \gamma'_{lt}|\mathbf{u}'_{lt}| & \dfrac{\gamma-1}{\beta^2}(\boldsymbol{\beta}'\cdot\hat{\mathbf{n}}')(\gamma'_{lt}|\mathbf{u}'_{lt}|) - \gamma(\gamma'_{lt}c) \\ n'_d|\mathbf{k}'| & \dfrac{\gamma-1}{\beta^2}(\boldsymbol{\beta}'\cdot\hat{\mathbf{n}}')(n'_d|\mathbf{k}'|) - \gamma\left(\dfrac{\omega'}{c}\right) \end{vmatrix} = 0\,. \tag{III-5}$$

Using $|\mathbf{u}'_{lt}| = c/n'_d$ and $|\mathbf{k}'| = \omega'/c$, we have

$$\frac{\dfrac{\gamma-1}{\beta^2}(\boldsymbol{\beta}'\cdot\hat{\mathbf{n}}')(\gamma'_{lt}|\mathbf{u}'_{lt}|) - \gamma(\gamma'_{lt}c)}{\gamma'_{lt}|\mathbf{u}'_{lt}|} = \frac{\gamma-1}{\beta^2}(\boldsymbol{\beta}'\cdot\hat{\mathbf{n}}') - n'_d\gamma\,, \tag{III-6}$$

$$\frac{\dfrac{\gamma-1}{\beta^2}(\boldsymbol{\beta}'\cdot\hat{\mathbf{n}}')(n'_d|\mathbf{k}'|) - \gamma\left(\dfrac{\omega'}{c}\right)}{n'_d|\mathbf{k}'|} = \frac{\gamma-1}{\beta^2}(\boldsymbol{\beta}'\cdot\hat{\mathbf{n}}') - \frac{\gamma}{n'_d}\,. \tag{III-7}$$

Since the sufficient and necessary condition for the holding of Eq. (III-5) is the holding of Eq. (III-6) = Eq. (III-7), we have

$$\frac{\gamma-1}{\beta^2}(\boldsymbol{\beta}'\cdot\hat{\mathbf{n}}') - n'_d\gamma = \frac{\gamma-1}{\beta^2}(\boldsymbol{\beta}'\cdot\hat{\mathbf{n}}') - \frac{\gamma}{n'_d}\,, \tag{III-8}$$

or

$$n'_d = 1\,. \tag{III-9}$$

From above it follows that, when $\hat{\mathbf{n}}'$ and $\boldsymbol{\beta}'$ are linearly independent, the sufficient and necessary condition for $\mathbf{u}_{lt} //(n_d\mathbf{k})$ is $n_d' = 1$ (vacuum).

Thus the above analysis can be outlined as follows:

(1) If $\boldsymbol{\beta}' // \hat{\mathbf{n}}'$ (including $\boldsymbol{\beta}' = 0$) is valid, then $\mathbf{u}_{lt} //(n_d\mathbf{k})$ holds.
(2) If $\hat{\mathbf{n}}'$ and $\boldsymbol{\beta}'$ are linearly independent, then $\mathbf{u}_{lt} //(n_d\mathbf{k})$ cannot hold except for in free space ($n_d' = 1$).

According to the principle of relativity, the light velocity must be parallel to the wave vector observed in any inertial frames. However the light particle velocity $\mathbf{u}_{lt}$ given by the 4-velocity $U^\mu = \gamma_{lt}(\mathbf{u}_{lt}, c)$ cannot meet the criterion. Therefore, the light 4-velocity in a moving medium, which follows Lorentz transformation as a Lorentz covariant first-rank tensor, is not compatible with the principle of relativity.

The light particle velocity can be expressed in terms of the medium moving velocity $\boldsymbol{\beta} = -\boldsymbol{\beta}'$ and the unit wave vector $\hat{\mathbf{n}}'$ in the medium-rest frame, given by

$$\mathbf{u}_{lt} = \left[\hat{\mathbf{n}}' + \left(\frac{\gamma^2}{\gamma+1}\hat{\mathbf{n}}'\cdot\boldsymbol{\beta} + \gamma n_d'\right)\boldsymbol{\beta}\right]\frac{c}{\gamma(n_d' + \hat{\mathbf{n}}'\cdot\boldsymbol{\beta})} \quad , \qquad \text{(light particle)}, \qquad \text{(III-10)}$$

while the light wave velocity (phase velocity) $\mathbf{v}_{ph} = \boldsymbol{\beta}_{ph}c$ defined by Eq. (10-1) in Sec. II is given by

$$\mathbf{v}_{ph} = \left[\hat{\mathbf{n}}' + \left(\frac{\gamma^2}{\gamma+1}\hat{\mathbf{n}}'\cdot\boldsymbol{\beta} + \frac{\gamma}{n_d'}\right)\boldsymbol{\beta}\right]\frac{n_d'\gamma(1+n_d'\hat{\mathbf{n}}'\cdot\boldsymbol{\beta})c}{(n_d'^2-1)+\gamma^2(1+n_d'\hat{\mathbf{n}}'\cdot\boldsymbol{\beta})^2}, \qquad \text{(light wave)}. \qquad \text{(III-11)}$$

When $|\boldsymbol{\beta}| << 1$, the above two velocity formulas can be approximated as

$$\mathbf{u}_{lt} \approx \frac{c}{n_d'}\left[1 - \frac{1}{n_d'}(\hat{\mathbf{n}}'\cdot\boldsymbol{\beta})\right]\hat{\mathbf{n}}' + \boldsymbol{\beta}c \quad , \qquad \text{(light particle)}, \qquad \text{(III-12)}$$

$$\mathbf{v}_{ph} \approx \frac{c}{n_d'}\left[1 + \left(n_d' - \frac{2}{n_d'}\right)(\hat{\mathbf{n}}'\cdot\boldsymbol{\beta})\right]\hat{\mathbf{n}}' + \frac{1}{n_d'^2}\boldsymbol{\beta}c\,, \qquad \text{(light wave)}. \qquad \text{(III-13)}$$

From above, we also can see that the results given by the two formulas are the same only when the medium moves parallel to the wave vector ($\boldsymbol{\beta} // \hat{\mathbf{n}}'$, Fizeau-experiment case) or in free space ($n_d' = 1$). Note: (i) $\mathbf{v}_{ph} //(n_d\mathbf{k})$ always holds including the $n_d' = 1$ (vacuum) case, according to the definition of phase velocity, Eq. (10-1) in Sec. II, and (ii) when $n_d' = 1$ (vacuum), $\gamma_{lt} \to \infty$ and $U^\mu = \gamma_{lt}(\mathbf{u}_{lt}, c)$ has no meaning mathematically, but $\mathbf{u}_{lt}$ given by Eq. (III-10) is still valid, which is a limit case.

**Remarks.** The basic requirements of the special principle of relativity on electrodynamics are (i) time-space coordinates $X^\mu = (\mathbf{x}, ct)$ follow Lorentz transformation, and (ii) Maxwell equations

$$\left(-\nabla\times\mathbf{E} - \frac{\partial(c\mathbf{B})}{\partial(c\mathbf{t})},\ \nabla\cdot(c\mathbf{B})\right) = (\mathbf{0}, 0)\,, \qquad \text{or} \qquad \partial_\mu \tilde{F}^{\mu\nu} = 0\,, \qquad \text{(III-14)}$$

$$\left(\nabla\times\mathbf{H} - \frac{\partial(c\mathbf{D})}{\partial(c\mathbf{t})},\ \nabla\cdot(c\mathbf{D})\right) = (\mathbf{J}, c\rho)\,, \qquad \text{or} \qquad \partial_\mu G^{\mu\nu} = J^\nu\,, \qquad \text{(III-15)}$$

are invariant in form under Lorentz transformations, where $\tilde{F}^{\mu\nu}$ and $G^{\mu\nu}$ are required to be second-rank Lorentz tensors, given by

$$\tilde{F}^{\mu\nu} = \begin{pmatrix} 0 & E_z & -E_y & cB_x \\ -E_z & 0 & E_x & cB_y \\ E_y & -E_x & 0 & cB_z \\ -cB_x & -cB_y & -cB_z & 0 \end{pmatrix}, \qquad \text{(III-16)}$$

$$G^{\mu\nu} = \begin{pmatrix} 0 & -H_z & H_y & D_x c \\ H_z & 0 & -H_x & D_y c \\ -H_y & H_x & 0 & D_z c \\ -D_x c & -D_y c & -D_z c & 0 \end{pmatrix}. \qquad \text{(III-17)}$$

We call the above two requirements *master physical laws* of electrodynamics, and call all others *sub-physical laws*. All sub-laws must be self-consistent with the master laws; for example, the Lorentz invariance of phase and the covariance of wave 4-vector are the direct results from the master laws, and they are intrinsically compatible with the principle of relativity. Required by the physical property of light momentum and the master physical laws, photon or wave phase velocity has two conditions to be imposed: (i) it should be parallel to the wave vector observed in any inertial frames (qualitative condition), and (ii) it should satisfy the equation of motion of equi-phase plane resulting from invariance of phase (quantitative condition). However the light particle 4-velocity Eq. (III-10) is generalized from the 4-velocity for massive particles, which is not required to be parallel to the wave vector; thus it is not surprising for the light particle 4-velocity not to be compatible with the principle of relativity. Since all sub-physical laws are required to be consistent with the master laws, sometimes it is not imperative to express the sub-laws in a Lorentz 4-vector or tensor form; for example, *the wave phase velocity defined by Eq. (10-1) or Eq. (10-2) in Sec. II is an invariant expression observed in all inertial frames, but it is not a tensor-form invariant expression.*

Since the wave 4-vector is given by $K^{\mu} = (n_d \mathbf{k},\ \omega/c) = \omega(n_d/c)^2(\boldsymbol{\beta}_{ph}c,\ c/n_d^2)$, the phase velocity $\boldsymbol{\beta}_{ph}c = (c/n_d)(\omega/|\omega|)\,\hat{\mathbf{n}}$ may be taken to be defined by the wave 4-vector if compared in form with the massive particle 4-velocity definition $\gamma_u(\mathbf{u}, c)$, with $\boldsymbol{\beta}_{ph}c$ as the "space component" of $K^{\mu}$ (although they are not the same dimension). Just because of this, Eq. (10-2) in Sec. II has an invariant form observed in all inertial frames.

Mathematically speaking, two linearly-related non-zero 4-vectors, which are parallel observed in any inertial frames, are different by a Lorentz invariant. However the factor $\omega(n_d/c)^2$ in $K^{\mu} = \omega(n_d/c)^2(\boldsymbol{\beta}_{ph}c,\ c/n_d^2)$ is not a Lorentz invariant, and $(\boldsymbol{\beta}_{ph}c,\ c/n_d^2)$ cannot be a "phase velocity 4-vector". In contrast, the photon 4-momentum $\hbar K^{\mu}$, which is parallel to $K^{\mu}$, is a 4-vector because the Planck constant $\hbar$ is a Lorentz invariant.

It should be emphasized that the phase-velocity definition Eq. (10-2) in Sec. II is invariant in form under the master laws of electrodynamics, but it is not an invariant expression in a tensor form. An invariant expression in a tensor form for the "phase velocity 4-vector" should have been defined in terms of $g_{\mu\nu}K^{\mu}\,dX^{\nu}/d\tau = g_{\mu\nu}K^{\mu}U^{\nu} = 0$ or $\gamma_{lt}(\omega - n_d \mathbf{k}\cdot\mathbf{u}_{lt}) = 0$ resulting from the invariance of phase $\Psi = (\omega t - n_d \mathbf{k}\cdot\mathbf{x}) = g_{\mu\nu}K^{\mu}X^{\nu}$, where the "phase velocity 4-vector" $U^{\mu} = \gamma_{lt}(\mathbf{u}_{lt}, c)$ is right the light particle 4-velocity given in Eqs. (III-1) and (III-2). Since $\mathbf{u}_{lt}\,//(n_d\mathbf{k})$ does not always hold in all frames as mentioned previously, *both* $U^{\mu} = \gamma_{lt}(\mathbf{u}_{lt}, c)$ *and* $g_{\mu\nu}K^{\mu}U^{\nu} = \gamma_{lt}(\omega - n_d\mathbf{k}\cdot\mathbf{u}_{lt}) = 0$ *are not compatible with the principle of relativity although they are both invariant expressions in a (first-rank) tensor form.*

From above discussions, we can conclude:

(1) A tensor form of physical law is no guarantee of its compatibility with the principle of relativity;

(2) All sub-physical laws must have invariant forms observed in all inertial frames and be compatible with the master physical laws, but the sub-physical laws do not have to be expressed in a tensor form.

**Photon's apparent velocity.** As mentioned above, it is not compatible with the principle of relativity if $U^{\mu} = \gamma_{lt}(\mathbf{u}_{lt}, c)$ is take as the "light particle (photon) 4-velocity" or "4-phase velocity", but analysis indicates that $U^{\mu} = \gamma_{lt}(\mathbf{u}_{lt}, c)$ is the photon's *apparent* 4-vector, which is shown below. (Note: $\mathbf{u}_{lt}$ is replaced by $\mathbf{u}$ in the following.)

Note: $\Psi = \omega t - n_d\mathbf{k}\cdot\mathbf{x} \Rightarrow \omega - n_d\mathbf{k}\cdot d\mathbf{x}/dt = 0$. Inserting the definitions $n_d\mathbf{k} \equiv |n_d\mathbf{k}|\,\hat{\mathbf{n}}$, $\boldsymbol{\beta}_{ph}c \equiv (\omega/|n_d\mathbf{k}|)\hat{\mathbf{n}}$, and $\mathbf{u} \equiv d\mathbf{x}/dt$, we obtain $\boldsymbol{\beta}_{ph}c = (\hat{\mathbf{n}}\cdot\mathbf{u})\hat{\mathbf{n}}$, where $\mathbf{u}$ is termed to be the photon's *apparent* velocity in the lab frame (confer Fig. 2 in Sec. II). In the medium-rest frame, $\mathbf{u}' = \boldsymbol{\beta}'_{ph}c = (c/n'_d)\hat{\mathbf{n}}'$ is assumed.

From Eq. (III-10), we have

$$\mathbf{u} = \left[\hat{\mathbf{n}}' + \left(\frac{\gamma - 1}{\beta^2}\hat{\mathbf{n}}'\cdot\boldsymbol{\beta}' - \gamma n'_d\right)\boldsymbol{\beta}'\right]\frac{c}{\gamma(n'_d - \hat{\mathbf{n}}'\cdot\boldsymbol{\beta}')} \quad . \qquad \text{(III-18)}$$

From Eq. (20) in Sec. IV, we have

$$\mathbf{E} = \gamma(1 - n'_d\hat{\mathbf{n}}'\cdot\boldsymbol{\beta}')\mathbf{E}' + \left(\gamma n'_d\hat{\mathbf{n}}' - \frac{\gamma - 1}{\beta^2}\boldsymbol{\beta}'\right)(\boldsymbol{\beta}'\cdot\mathbf{E}'), \qquad \text{(III-19)}$$

$$\mathbf{H} = \gamma(1 - n'_d\hat{\mathbf{n}}'\cdot\boldsymbol{\beta}')\mathbf{H}' + \left(\gamma n'_d\hat{\mathbf{n}}' - \frac{\gamma - 1}{\beta^2}\boldsymbol{\beta}'\right)(\boldsymbol{\beta}'\cdot\mathbf{H}'). \qquad \text{(III-20)}$$

From above, we obtain

$$\mathbf{E}\times\mathbf{H}=\gamma(1-n_d'\hat{\mathbf{n}}'\cdot\boldsymbol{\beta}')\left[\hat{\mathbf{n}}'+\left(\frac{\gamma-1}{\beta^2}(\hat{\mathbf{n}}'\cdot\boldsymbol{\beta}')-\gamma n_d'\right)\boldsymbol{\beta}'\right]\left(\frac{c}{n_d'}\right)(\mathbf{D}'\cdot\mathbf{E}')\,. \tag{III-21}$$

From Eq. (31) in Sec. IV, we have

$$\mathbf{D}\cdot\mathbf{E}=\gamma^2(1-n_d'\hat{\mathbf{n}}'\cdot\boldsymbol{\beta}')\left(1-\frac{1}{n_d'}\hat{\mathbf{n}}'\cdot\boldsymbol{\beta}'\right)\mathbf{D}'\cdot\mathbf{E}'\,. \tag{III-22}$$

From above Eqs. (III-21) and (III-22), with $\mathbf{B}\cdot\mathbf{H}=\mathbf{D}\cdot\mathbf{E}$ considered, we have

$$\frac{\mathbf{E}\times\mathbf{H}}{0.5(\mathbf{D}\cdot\mathbf{E}+\mathbf{B}\cdot\mathbf{H})}=\frac{\mathbf{E}\times\mathbf{H}}{\mathbf{D}\cdot\mathbf{E}}=\left[\hat{\mathbf{n}}'+\left(\frac{\gamma-1}{\beta^2}(\hat{\mathbf{n}}'\cdot\boldsymbol{\beta}')-\gamma n_d'\right)\boldsymbol{\beta}'\right]\frac{c}{\gamma(n_d'-\hat{\mathbf{n}}'\cdot\boldsymbol{\beta}')}\,. \tag{III-23}$$

Comparing above Eq. (III-23) with Eq. (III-18), we have

$$\mathbf{u}=\frac{\mathbf{E}\times\mathbf{H}}{0.5(\mathbf{D}\cdot\mathbf{E}+\mathbf{B}\cdot\mathbf{H})}=\frac{\mathbf{S}}{W_{em}}\,, \tag{III-24}$$

namely, the photon's apparent velocity is equal to the "energy velocity $\mathbf{S}/W_{em}$".

## Attachment-IV. A mathematical example

Some scientists firmly claim that

> "…**it is a well known result that even if $T_\mu^{\ 0}$ is not a vector, its integral over space (in each inertial frame) $\int T_\mu^{\ 0} d^3x$ is indeed a Lorentz vector**. This is a consequence of the transformation of the quantities (and of the spacelike region where the integral is defined) under Lorentz transformations. The same happens with the integral $\int \rho d^3x$ corresponding to the total electric charge. The total charge is a scalar even when $\rho$ is "only the 0-th component" of the 4-current density $J^\mu=(c\rho,\mathbf{J})$. This is all well known, as can be read for instance in Jackson's book [J.D. Jackson, Classical Electrodynamics, (John Wiley & Sons, NJ, 1999) 3rd Ed.], page 555. For more detail see also [S. Weinberg, Gravitation and Cosmology, (Wiley, NY, 1972)], pages 40-41, or [R. U. Sexl and H. K. Urbantke, Relativity, Groups, Particles, (Springer-Verlag Wien, NY, 1992)], pages 106-107."

First, it should be indicated that there is no such "a well known result" in the reference books cited above. In this attachment, a specific math example is constructed to show the above claim is not correct.

Suppose that $X'Y'Z'$ frame moves at $\beta c$ with respect to the lab frame $XYZ$ in the $x$-direction. The time-space Lorentz transformation is given by

$$x'=\gamma(x-\beta ct)\,,\quad y'=y\,,\quad z'=z\,,\quad ct'=\gamma(ct-\beta x)\,. \tag{IV-1}$$

We label $x'=X'^1$, $y'=X'^2$, $z'=X'^3$, and $ct'=X'^4$, and we have

$$\begin{pmatrix}X'^1\\X'^2\\X'^3\\X'^4\end{pmatrix}=\begin{pmatrix}\gamma&0&0&-\beta\gamma\\0&1&0&0\\0&0&1&0\\-\beta\gamma&0&0&\gamma\end{pmatrix}\begin{pmatrix}X^1\\X^2\\X^3\\X^4\end{pmatrix},\qquad\text{or}\qquad\begin{pmatrix}X^1\\X^2\\X^3\\X^4\end{pmatrix}=\begin{pmatrix}\gamma&0&0&\beta\gamma\\0&1&0&0\\0&0&1&0\\\beta\gamma&0&0&\gamma\end{pmatrix}\begin{pmatrix}X'^1\\X'^2\\X'^3\\X'^4\end{pmatrix}. \tag{IV-2}$$

The metric tensor is given by $g^{\mu\nu}=g_{\mu\nu}=diag(-1,-1,-1,+1)$, and $g^{\mu\nu}=g^{\mu\lambda}g_\lambda^{\ \nu}$ with $g_\mu^{\ \nu}=diag(1,1,1,1)$. Obviously, $g_\mu^{\ \nu}=g_{\mu\lambda}g^{\lambda\nu}$ is also a tensor.

If $A'^\mu$ and $A'_\mu$, and $A^\mu$ and $A_\mu$ are, respectively, contra- and co-variant expressions in $X'Y'Z'$ and $XYZ$ frames for an arbitrary 4-vector, then their Lorentz transformations are given by

$$\begin{pmatrix} A'^1 \\ A'^2 \\ A'^3 \\ A'^4 \end{pmatrix} = \begin{pmatrix} \gamma & 0 & 0 & -\beta\gamma \\ 0 & 1 & 0 & 0 \\ 0 & 0 & 1 & 0 \\ -\beta\gamma & 0 & 0 & \gamma \end{pmatrix} \begin{pmatrix} A^1 \\ A^2 \\ A^3 \\ A^4 \end{pmatrix}, \qquad \begin{pmatrix} A'_1 \\ A'_2 \\ A'_3 \\ A'_4 \end{pmatrix} = \begin{pmatrix} \gamma & 0 & 0 & \beta\gamma \\ 0 & 1 & 0 & 0 \\ 0 & 0 & 1 & 0 \\ \beta\gamma & 0 & 0 & \gamma \end{pmatrix} \begin{pmatrix} A_1 \\ A_2 \\ A_3 \\ A_3 \end{pmatrix}. \tag{IV-3}$$

Now let us construct a symmetric second-rank tensor, given by in $X'Y'Z'$ frame

$$T'^{\mu\nu} = g^{\mu\nu}\delta(\mathbf{x}') \text{, with } \delta(\mathbf{x}') = \delta(x')\delta(y')\delta(z') \text{ the delta-function,} \tag{IV-4}$$

and its co-contra-variant expression is given by

$$T'^{\nu}_{\mu} = g_{\mu}^{\ \nu}\delta(\mathbf{x}') . \tag{IV-5}$$

We define

$$P'_{\mu} = \iiint T'^{4}_{\mu} dV' = \iiint g_{\mu}^{\ 4}\delta(\mathbf{x}')dV' = g_{\mu}^{\ 4} \iiint_{\text{all space}} \delta(\mathbf{x}')dx'dy'dz' = \begin{pmatrix} 0 \\ 0 \\ 0 \\ 1 \end{pmatrix}. \tag{IV-6}$$

Note: $T'^{0}_{\mu}$ in the Letter by Ramos, Rubilar, and Obukhov [Phys. Lett. A **375**, 1703 (2011)] corresponds to $T'^{4}_{\mu}$ here.

Obviously, $g^{\mu\nu}$ and $g_{\mu}^{\ \nu}$ are the same under Lorentz transformations, and we have the corresponding expressions in the $XYZ$ frame, given by

$$T^{\mu\nu} = g^{\mu\nu}\delta(\mathbf{x}') , \qquad T^{\nu}_{\mu} = g_{\mu}^{\ \nu}\delta(\mathbf{x}') , \tag{IV-7}$$

where

$$\delta(\mathbf{x}') = \delta[\gamma(x-\beta ct)]\delta(y)\delta(z) = \frac{1}{\gamma}\delta(x-\beta ct)\delta(y)\delta(z) . \tag{IV-8}$$

Thus we have

$$P_{\mu} = \iiint T^{4}_{\mu} dV = \iiint g_{\mu}^{\ 4}\delta(\mathbf{x}')dxdydz = g_{\mu}^{\ 4} \iiint_{\text{all space}} \frac{1}{\gamma}\delta(x-\beta ct)\delta(y)\delta(z)dxdydz = \frac{1}{\gamma}\begin{pmatrix} 0 \\ 0 \\ 0 \\ 1 \end{pmatrix}. \tag{IV-9}$$

Now let us check if $P_{\mu}$ is a Lorentz covariant 4-vector by checking the relation between $P'_{\mu}$ and $P_{\mu}$. From Eq. (IV-6) and Eq. (IV-9), we have

$$P'_{\mu} = \gamma P_{\mu} , \qquad \text{or} \qquad P'_{\mu} = \gamma g_{\mu}^{\ \nu} P_{\nu} . \tag{IV-10}$$

Obviously, $\gamma g_{\mu}^{\ \nu}$ is not equal to the Lorentz transformation given in Eq. (IV-3) unless $\beta = 0$ ($\gamma = 1$).

**Therefore, we can conclude that there is no such a general math rule for tensors:**

"…even if $T_{\mu}^{\ 0}$ is not a vector, its integral over space (in each inertial frame) $\int T_{\mu}^{\ 0} d^3x$ is indeed a Lorentz vector."

**First-rank tensor example.** One well-known example is given below to support our above conclusion.

Suppose that $X'Y'Z'$ frame moves uniformly at $\boldsymbol{\beta}c$ with respect to $XYZ$ frame, and the Lorentz transformation is given by $x' = \gamma(x-\beta ct)$ , $y' = y$ , $z' = z$ , and $ct' = \gamma(ct-\beta x)$ , with the metric given by $g^{\mu\nu} = g_{\mu\nu} = diag(-1,-1,-1,+1)$ and the transformation of Dirac function given by $\delta(\mathbf{x}') = \gamma^{-1}\delta(\mathbf{x}-\boldsymbol{\beta}ct)$. A unit point charge with a density of $\rho' = \delta(\mathbf{x}')$ rests at $\mathbf{x}' = 0$, and the current density 4-vector is given by

$$J'^{\mu} = (\mathbf{0}, c\delta(\mathbf{x}')) , \qquad \text{with} \qquad g_{\mu\nu}J'^{\mu}J'^{\nu} = c^2\delta^2(\mathbf{x}') , \tag{IV-11}$$

and we have

$$P'^{\mu} = \int J'^{\mu} dV' = \int J'^{\mu} dx'dy'dz' = (\mathbf{0}, c) , \qquad \text{with} \qquad g_{\mu\nu} P'^{\mu} P'^{\nu} = c^2 . \tag{IV-12}$$

Observed in $XYZ$ frame, we have

$$J^{\mu} = (\, \gamma\boldsymbol{\beta} c\delta(\mathbf{x}'), \gamma c\delta(\mathbf{x}') \,) , \qquad \text{with} \qquad g_{\mu\nu} J^{\mu} J^{\nu} = c^2 \delta^2(\mathbf{x}') = g_{\mu\nu} J'^{\mu} J'^{\nu} , \tag{IV-13}$$

$$P^{\mu} = \int J^{\mu} dV = (\, \boldsymbol{\beta} c, c \,) , \qquad \text{with} \qquad g_{\mu\nu} P^{\mu} P^{\nu} = c^2 (1-\beta^2) \neq g_{\mu\nu} P'^{\mu} P'^{\nu} . \tag{IV-14}$$

From above it is seen that, although $J^{\mu}$ is a Lorentz covariant 4-vector or a first-rank tensor, after its integration over the space, $\int J^{\mu} dV$ is not Lorentz covariant any more, because its "4D-length square", $g_{\mu\nu} P^{\mu} P^{\nu}$ does not keep invariant under the Lorentz transformation.

**Can the Lorentz invariance of total charge be taken as a general math rule for tensors?** According to the textbooks [S. Weinberg, Gravitation and Cosmology, (Wiley, NY, 1972), pages 40-41; J. D. Jackson, Classical Electrodynamics, (John Wiley & Sons, NJ, 1999), 3rd Ed., page 555], the current density satisfies the continuity equation $\nabla \cdot \mathbf{J} + \partial\rho/\partial t = 0$ , or $J^{\mu} = (\mathbf{J}, c\rho)$ satisfies the invariant conservation equation $\partial_{\mu} J^{\mu} = 0$ in the Minkowski space with the metric $g^{\mu\nu} = g_{\mu\nu} =$ diag(-1,-1,-1,+1); $cQ = \int J^4 d^3x = \int c\rho d^3x$ is a Lorentz invariant.

The above-mentioned Weinberg's book has a specific statement on p. 40:

> Whenever **any current** $J^{\mu} = (\mathbf{J}, c\rho)$ satisfies the invariant conservation law $\partial_{\mu} J^{\mu} = 0$ , we can form a total charge $Q = \int (J^4/c) d^3x = \int \rho d^3x$ (which is a Lorentz invariant).

From the above statement, one might conjecture that a "general conclusion" can be obtained:

> Whenever **any 4-vector** $A^{\mu} = (\mathbf{A}, A^4)$ satisfies the invariant conservation law $\partial_{\mu} A^{\mu} = 0$, we can form a quantity $\int (A^4) d^3x$ , which is a Lorentz invariant.

Unfortunately, this conjectured "general conclusion" is not correct, because the current density $J^{\mu} = (\mathbf{J}, c\rho)$ is a special kind of 4-vector, instead of a general 4-vector. For a current density, in addition to $\partial_{\mu} J^{\mu} = 0$ , a necessary condition $|\mathbf{J}/c\rho| < 1$ must hold, for example. Below, a specific math example is constructed to illustrate that the above conjectured "general conclusion" is indeed not correct.

Suppose that $X'Y'Z'$ frame moves at $\beta c$ with respect to the lab frame $XYZ$ along the *x*-direction. Observed in the moving frame $X'Y'Z'$ , we define a 4-vector $A'^{\mu} = (\mathbf{A}', A'^4)$ , with $A'_x = const > 0$ , $A'_y = A'_z = 0$ and $A'^4 = 0$ , and $|\mathbf{A}'/A'^4| = +\infty > 1$ , violating the current-density condition $|\mathbf{J}/c\rho| < 1$ . Note that $\partial'_{\mu} A'^{\mu} = 0$ is satisfied. We have $\int A'^4 d^3x' = 0$ .

Observed in the lab frame $XYZ$ , from the Lorentz transformation we have $A^{\mu} = (\mathbf{A}, A^4)$ , with $A_x = \gamma A'_x = const > 0$ , $A_y = A_z = 0$ , and $A^4 = \gamma\beta A'_x = const > 0$ , and $|\mathbf{A}/A^4| = |1/\beta| > 1$ , keeping violating the current-density condition $|\mathbf{J}/c\rho| < 1$ . Note that $\partial_{\mu} A^{\mu} = 0$ still holds, but $\int A^4 d^3x = \int (\gamma\beta A'_x) d^3x = \infty$ . Thus we have $\int A^4 d^3x \neq \int A'^4 d^3x'$ , which is not a Lorentz invariant.

Therefore, the Lorentz invariance of total charge cannot be taken as a general math rule for tensors.